\documentclass[prb,twocolumn,superscriptaddress,showpacs]{revtex4}
\usepackage{mathrsfs}
\usepackage{amsmath,amsfonts,amssymb}
\usepackage{epsfig}
\usepackage{graphicx}
\usepackage{wasysym}
\usepackage{bbm}
\usepackage{psfrag}
\usepackage{color}
\usepackage{pstool}
\usepackage{braket}
\usepackage{lmodern}
\usepackage[dvips, bookmarks, colorlinks=true, plainpages = false, citecolor = blue, linkcolor = blue, urlcolor = blue, filecolor = blue]{hyperref}
\def\be{\begin{equation}}
\def\ee{\end{equation}}
\def\bea{\begin{eqnarray}}
\def\eea{\end{eqnarray}}

\begin{document}
\title{Topological magnons in ferromagnetic Kitaev-Heisenberg model on CaVO lattice}

\author{Moumita Deb}\email{moumitadeb44@gmail.com}
\author{Asim Kumar Ghosh}
 \email{asimkumar96@yahoo.com}
\affiliation {Department of Physics, Jadavpur University, 
188 Raja Subodh Chandra Mallik Road, Kolkata 700032, India}
\begin{abstract}
A number of topological phases are found to emerge in the ferromagnetic Kitaev-Heisenberg model
on CaVO lattice in the presence of Dzyaloshinskii-Moriya interaction. Heisenberg and Kitaev terms
have been considered on nearest and next-nearest neighbor bonds in a variety of ways. Both isotropic
and anisotropic couplings are taken into account. Topological phases are characterized by Chern
numbers for the distinct magnon bands as well as the number of modes for topologically protected
gapless magnon edge states. Band structure, dispersion relation along the high-symmetric points of
first Brillouin zone, density of states and thermal Hall conductance have been evaluated for every
phase. An extensive Phase diagram has been constructed. Topological phase transition in the
parameter space is also studied.
\end{abstract}
\maketitle
\section{INTRODUCTION}
Study of topological properties on condensed matter systems 
hails a new era in theoretical as well as experimental
investigations. 
Chern number $({\mathcal C})$ is recognized as the 
topological invariant for the characterization of topological phase 
of that particular class of topological insulators (TI), 
where time-reversal symmetry (TRS) is broken \cite{TKNN}. 
The relation between  ${\mathcal C}$ and number of topologically protected
modes of edge state is governed by 
`bulk-edge-correspondence' rule \cite{Hatsugai}. 
${\mathcal C}$ of a definite band is determined by integrating 
the Berry curvature over the first Brillouin zone (BZ).
Nontrivial topological phase corresponds to that 
state where the system exhibits at least a pair of 
nonzero values of ${\mathcal C}$. 
A variety of many particle interactions are found 
responsible behind the emergence of non-trivial topological phases. 
Several kinds of interactions deform the Berry curvatures 
in so many different ways that they eventually lead to 
numerous topological phases. Performance of few potential 
two-spin interactions will be discussed here, those are 
found crucial to induce non-triviality  
in magnetic systems. 

The first magnetic material who demonstrates non-triviality 
is Lu$_2$V$_2$O$_7$ \cite{Tokura1,Lee}. The experimental results 
are explained in terms of a spin-1/2 ferromagnetic (FM) Heisenberg 
model on a kagom\'e lattice in the presence of 
antisymmetric Dzyaloshinskii-Moriya 
interaction (DMI) \cite{Dzyaloshinskii,Moriya}. 
Two distinct topological phases characterized by 
${\mathcal C}$=($10\bar{1}$) and ($\bar 101$)
appear around the zero DMI strength,    
where $\bar x$ means $-x$. Emergence of those  
phases can be regarded as the handiwork of DMI term \cite{Li,Sen}. 
Those states are observed again in another kagom\'e ferromagnet, 
Cu[1,3-benzenedicarboxylate (bdc)] \cite{Chisnell}.  
This kind of state is now termed as 
topological magnon insulating (TMI) phase. 
In both cases, TMI phase is observed 
in the presence of an external magnetic field 
which acts as a TRS breaking component in the system. 

In another development, FM Heisenberg models consisting of 
both nearest-neighbor (NN) Kitaev \cite{Kitaev} and symmetric 
spin-anisotropic interactions (SAI) 
are found to exhibit TMI phases on the honeycomb lattice 
in the presence of magnetic field. This two-band system hosts 
the TMI phase with ${\mathcal C}$=($1\bar 1$), when both the 
two-spin interacting terms are present \cite{Joshi}. 
This model is extended beyond NN interactions by including 
third-neighbor Kitaev and SAI terms 
which is found to host multiple novel TMI phases with higher 
Chern numbers \cite{Moumita}. DMI term fails to induce 
non-triviality in this honeycomb model. 
It is
worth mentioning that multiple TMI phases are found in
the triplet six-spin plaquette excitations of the antifer-
romagnetic (AFM) Heisenberg model on the honeycomb
lattice without the DMI terms \cite{Moumita_2}.
Thus search for new combinations of two-spin terms 
continues those are capable to induce non-triviality in other 
lattice geometries. 

In this article, emergence of multiple TMI phases 
will be reported in a four-band FM Heisenberg model formulated 
on a CaVO lattice which includes both 
NN and next-nearest-neighbor (NNN) Heisenberg and Kitaev 
terms along with NNN DMI term. 
NN DMI term has no effect on the topological phases, while presence of NNN DMI turnes out to be indispensable for the emergence of 
topological phases. 
Again, DMI only on the NN bonds alone fails to 
drive the system into the topologically nontrivial regime. 
Remarkably, SAI is found to play in this system while 
the TRS breaking magnetic field
is taken into account for obvious reason. Emergence of
photo-induced multiple topological phases is reported be-
fore in a tight-binding model on CaVO lattice \cite{Sil}. Exis-
tence of multiple Dirac nodal-lines in AFM magnons and
triplet excitations are noted for a Heisenberg model on
this lattice \cite{Owerre,Moumita_3}.

CaVO lattice has been brought to light 
before in the context of AFM compound, 
CaV$_4$O$_9$ \cite{Taniguchi}. 
The spin-1/2 V$^{4+}$ ions in this frustrated system 
constitute  CaVO lattice structure \cite{Indrani}. 
Coordination number for both honeycomb and CaVO lattices 
is three while their symmetries are different. 
For example, honeycomb (CaVO) lattice 
has the six (four) -fold rotational symmetry,  $C_6$ ($C_4$). 
Primitive cell contains two and four lattice points 
for honeycomb and  CaVO lattices, respectively. 
FM Kitaev models with anisotropic NN bond strengths based 
on those non-Bravais lattices have been solved exactly 
in terms of Majorana fermions \cite{Kitaev,Sun}. 
The gapless phases for both the lattices become topologically 
nontrivial as soon as the magnetic field is switched on. 
The topological phase of both honeycomb and CaVO Kitaev models 
is unique in a sense that the resulting two-band 
system carries ${\mathcal C}$=($1\bar 1$) 
in the Majorana fermion representation.  
In contrast, no topological phase based on the 
bosonic magnon excitation emerges 
in the NN Kitav models on honeycomb and CaVO lattices 
in the presence of magnetic field.  

  \begin{figure*}[t]
\psfrag{1}{ (a)}
\psfrag{2}{ (c)}
\psfrag{3}{ (b)}
\psfrag{A}{ $A$}
\psfrag{B}{ $B$}
\psfrag{C}{ $C$}
\psfrag{D}{ $D$}
\psfrag{a1}{ $\delta_1$}
\psfrag{a2}{ $\delta_2$}
\psfrag{aa}{ $a$}
\psfrag{x}{ $\hat{x}$}
\psfrag{y}{ $\hat{y}$}
\psfrag{a}{ $K_x$}
\psfrag{b}{ $K_y$}
\psfrag{c}{ $K_z$}
\psfrag{d}{ $K^\prime_x$}
\psfrag{e}{ $K^\prime_y$}
\psfrag{f}{ $K^\prime_z$}
\psfrag{g}{ $D_{\rm m}$}
\psfrag{N}{ $N$ units}
\psfrag{K1}{ $k_x$}
\psfrag{K2}{ $k_y$}
\psfrag{G}{  $\Gamma$}
\psfrag{X}{ X}
\psfrag{M}{ M}
   \centering
   \includegraphics[width=350pt]{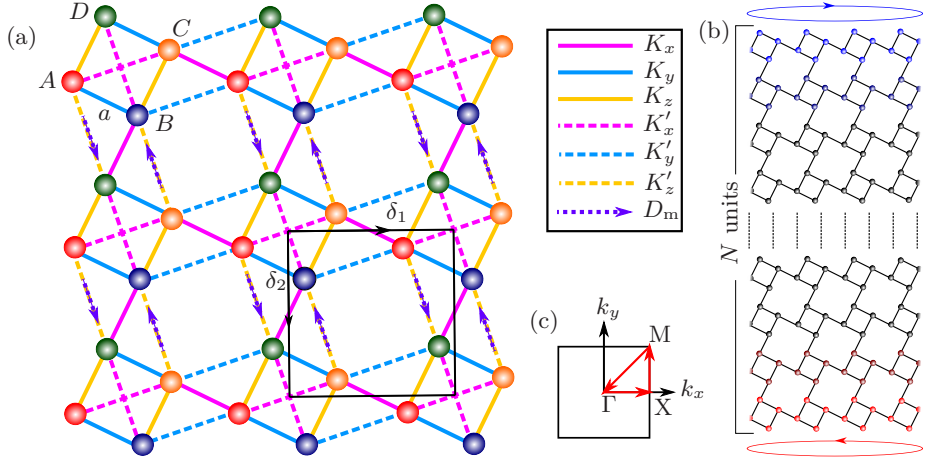}
  \caption{(a) Geometrical view of the model with NN and NNN interactions 
on the CaVO lattice. Opposite arrowheads indicate
the opposite directions of DMI terms. (b) Geometry of the lattice used for edge 
state calculation, upper and lower edges are drawn in blue and red colors, 
respectively. (c) The first Brillouin zone showing the high-symmetry 
points, $\Gamma$, M and X.}
   \label{lattice}
     \end{figure*}  

Hamiltonian is formulated in 
Sec \ref{KH}. The linear spin-wave theory (LSWT) is 
developed here. Numerical 
evaluation of ${\mathcal C}$, edge states and 
thermal Hall conductance (THC) have been described in 
Sec \ref{CETHC}. THC experiences sudden jump in the vicinity 
of phase transition points. 
The system exhibits seventeen distinct topological phases
in total. 
Half dozen of them appears when isotropic 
Kitaev interaction is assumed while a dozen appears for anisotropic 
case. Only one phase is found common in both the cases. 
Topological phase diagrams have been produced in various forms. 
A discussion based on those results is available 
in Sec \ref{Discussion}.
\section{Kitaev-Heisenberg model with
Dzyaloshinskii-Moriya interactions on the CaVO lattice }
\label{KH}
In order to develop the spin wave theory, a general form of spin Hamiltonian is 
considered which contains three specific two-spin interacting terms, say, 
Kitaev, Heisenberg and DMI in the presence of 
an external magnetic field. Both Kitaev and Heisenberg terms are present 
on the NN and NNN bonds 
of the CaVO lattice while DMI is present on a pair of opposite 
NNN bonds within the octagon plaquette. 
 The Hamiltonian (Eq \ref{ham}) is written as
\begin{equation}
 \begin{aligned}
{\mathcal H}&=J\sum\limits_{\langle ij \rangle} \boldsymbol{S}_i\cdot\boldsymbol{S}_j
+2\sum\limits_{\langle ij \rangle_\gamma}\!K_\gamma \;S^\gamma_i S^\gamma_j
+J'\!\sum\limits_{\langle\langle ij \rangle\rangle} \!\boldsymbol{S}_i\cdot\boldsymbol{S}_j \\
& +2\!\sum\limits_{\langle\langle ij \rangle\rangle_\gamma}\!K'_\gamma \,S^\gamma_i S^\gamma_j
 \!+\boldsymbol{D}_{\rm m}\cdot\!\sum\limits_{\langle\langle ij \rangle\rangle} \!\boldsymbol{S}_i\times\boldsymbol{S}_j
-\boldsymbol{h}\cdot\sum\limits_{i}  \boldsymbol{S}_i.
 \label{ham}
\end{aligned}
  \end{equation}
  
Here, $J$ ($J^\prime$) and $K_\gamma$ ($K^\prime_\gamma$) are the Heisenberg and Kitaev interaction 
strengths respectively for the NN (NNN) bonds. DMI strength is denoted by $D_{\rm m}$. 
Every site of the CaVO lattice is connected 
with the others by three NN bonds as well as three NNN bonds.
In order to assign the Kitaev interactions, three different 
components are represented by $\gamma=x,y,z$, both for NN and NNN bonds. 
${\boldsymbol h}=g\mu_{\rm B}{\boldsymbol H}$, 
where $H$ is the strength of magnetic field which is 
acting along the $+\hat z$ direction. 
$S^\alpha_i$ is the $\alpha$-th component of spin operator, 
${\boldsymbol S}_i$, at the $i$-th site, where $\alpha=x,y,z$.  
Summations over NN and NNN  
bonds are shown by the indices  ${\langle \cdot \rangle}$ and 
${\langle\langle \cdot \rangle\rangle} $, respectively. 
Periodic boundary condition (PBC) is assumed along both $x$ and $y$ 
directions. Values of $J$ and $J'$ are always negative when
they are nonzero. 
Both positive and negative values of $K$ and $K'$ are considered, 
but magnitudes of $J$ and $J'$ are always greater than  
those of $K$ and $K'$ in the isotropic case to ensure the FM ground state. 
$J$ and $J'$ are assumed zero in the anisotropic case, without any loss of
generality.

Schematic view of this spin model is given in Fig \ref{lattice} (a), where  
NN (NNN) bonds are indicated by solid (dashed) lines.
${\bf \delta_1}$ and ${\bf \delta_2}$ are the two primitive vectors 
to constitute the primitive cell (a square of arm length $\sqrt 5 \,a$) 
of the CaVO lattice. 
The primitive cell represented by the square encloses four nonequivalent sites 
$A$, $B$, $C$ and $D$, shown by red, blue, yellow and green spheres, respectively. 
Obviously, the CaVO lattice can be decomposed in terms of four 
interpenetrating square lattices made of each for four sites 
$A$, $B$, $C$ and $D$, separately. 
Alternately, the resulting lattice can be thought of as a 
specific structure of 1/5-depleted square lattice which preserves the 
four-fold rotational symmetry, ${\mathcal C}_4$, of the square lattice itself. 

No topological phase will appear in this model if DMI is absent.
Additionally, the direction of $\boldsymbol{D}_{\rm m}$ as well as the combination of 
bonds on which DMI is acting are very crucial for the emergence 
of topological phases. For example, DMI on every NN and NNN bond within the 
square plaquette does not lead to non-trivial topological phase 
by any means. 
It is found that only two specific combinations comprising of 
two pairs of opposite NNN bonds within the octagon plaquette 
where the directions of $\boldsymbol{D}_{\rm m}$ are also opposite 
to each other could lead to the nontriviality. 
One of such combination is shown in Fig \ref{lattice} (a), in which 
DMI is acting over $AD$ and $BC$ bonds but opposite in directions. 
Opposite arrowhead over  $AD$ and $BC$ bonds imply the directions of 
$\boldsymbol{D}_{\rm m}$, which are $\pm \hat z$, for the respective bonds. 
DMI over $AB$ and $CD$ bonds may form another potential combination 
if the directions of that are chosen opposite to each other.  
However, the later choice is not assumed here, 
since it fails to exhibit
additional topological phases by any means. 

  
\begin{figure}[t]
 \psfrag{E}{ $E$}
 \psfrag{kx}{\tiny $k_x$}
 \psfrag{ky}{\tiny $k_y$}
   \psfrag{AA}{(a)}
   \psfrag{c1}{\tiny ${\mathcal C}_1=1$}
\psfrag{c2}{\tiny ${\mathcal C}_2=1$}
\psfrag{c3}{\tiny ${\mathcal C}_3=\bar 1$}
\psfrag{c4}{\tiny ${\mathcal C}_4=\bar 1$}
\begin{minipage}{0.18\textwidth}
  \includegraphics[width=170pt]{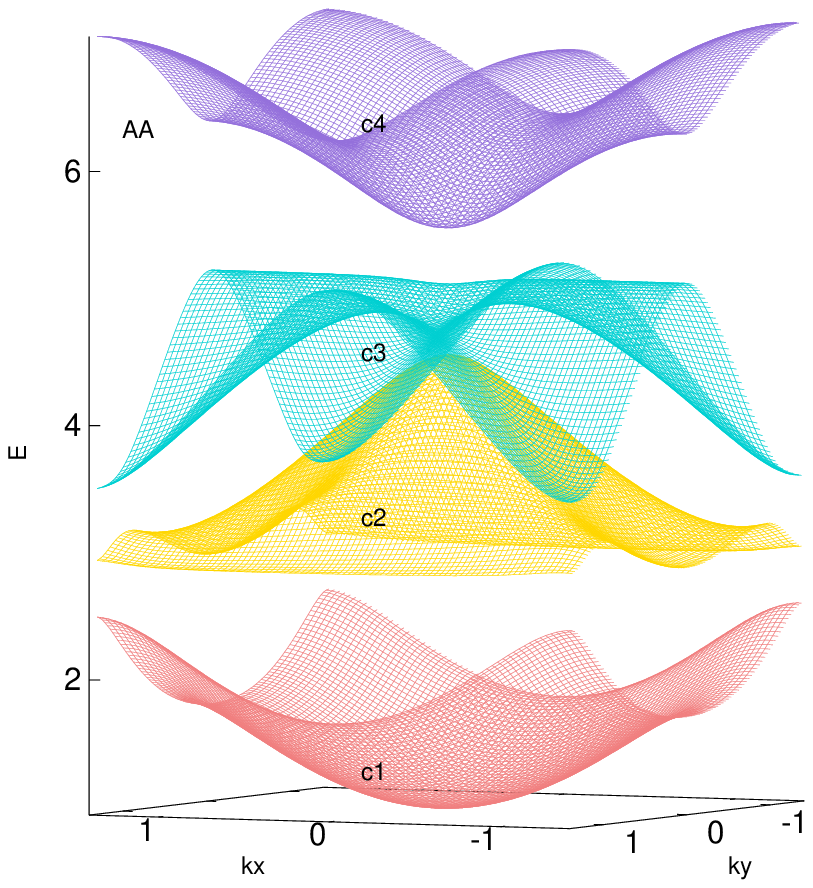}
  \end{minipage}\hfill
   \begin{minipage}{0.28\textwidth}
  \psfrag{E}{}
  \psfrag{BB}{(b)}
    \psfrag{c1}{\tiny ${\mathcal C}_1=1$}
\psfrag{c2}{\tiny ${\mathcal C}_2=1$}
\psfrag{c3}{\tiny ${\mathcal C}_3=\bar 2$}
\psfrag{c4}{\tiny ${\mathcal C}_4=0$}
    \centering
  \includegraphics[width=170pt]{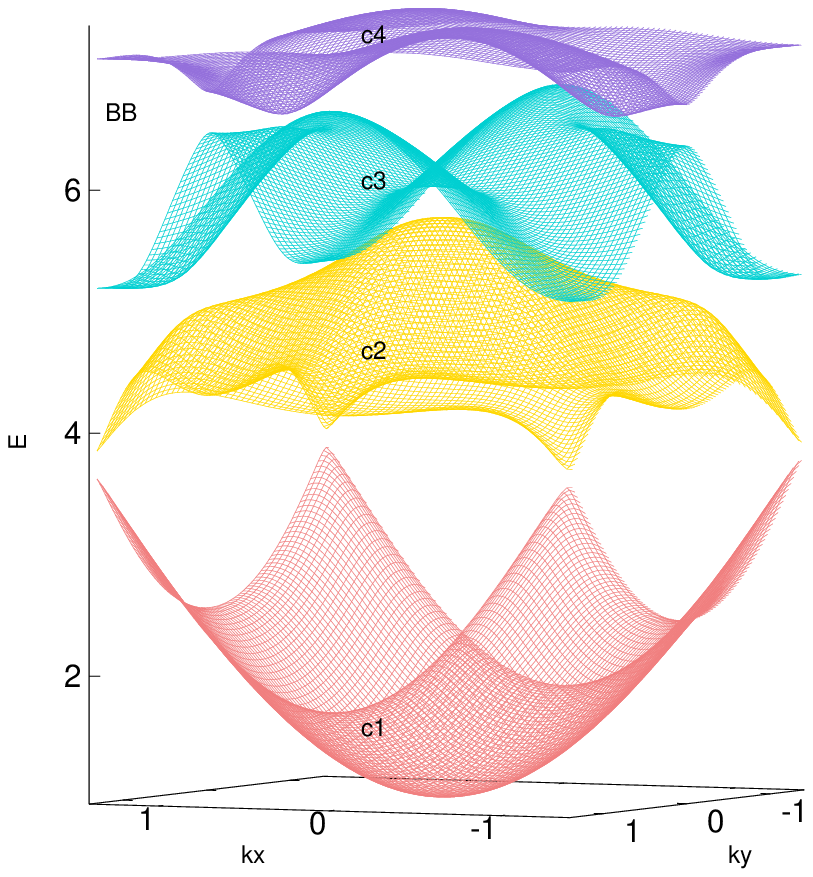}
   \end{minipage}\hfill 
  \psfrag{E}{E}
\psfrag{CC}{(c)}
   \psfrag{c1}{\tiny ${\mathcal C}_1=1$}
\psfrag{c2}{\tiny ${\mathcal C}_2=\bar 1$}
\psfrag{c3}{\tiny ${\mathcal C}_3=1$}
\psfrag{c4}{\tiny ${\mathcal C}_4=\bar 1$}
    \centering
    \begin{minipage}{0.18\textwidth}
  \includegraphics[width=170pt]{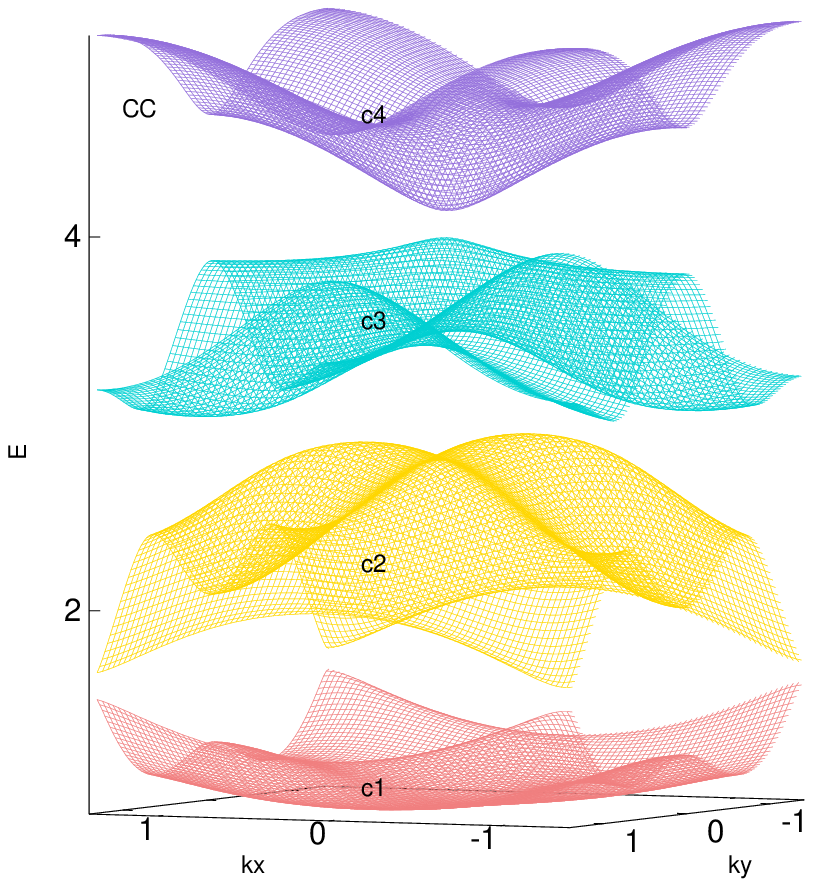}
  \end{minipage}\hfill
   \begin{minipage}{0.28\textwidth}
\psfrag{E}{}
\psfrag{DD}{(d)}
    \psfrag{c1}{\tiny ${\mathcal C}_1=0$}
\psfrag{c2}{\tiny ${\mathcal C}_2=0$}
\psfrag{c3}{\tiny ${\mathcal C}_3=2$}
\psfrag{c4}{\tiny ${\mathcal C}_4=\bar 2$}
    \centering
  \includegraphics[width=170pt]{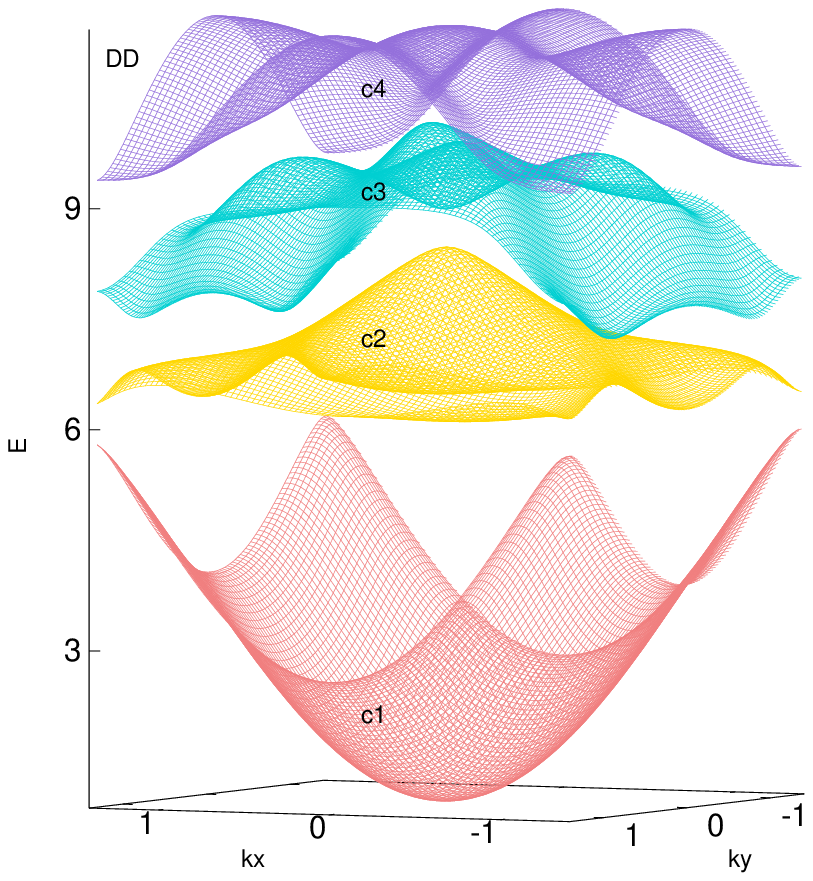}
   \end{minipage}\hfill
\psfrag{E}{E}
 \psfrag{kx}{\tiny $k_x$}
 \psfrag{ky}{\tiny $k_y$}
   \psfrag{EE}{(e)}
   \psfrag{c1}{\tiny $C_1=\bar 2$}
\psfrag{c2}{\tiny ${\mathcal C}_2=2$}
\psfrag{c3}{\tiny ${\mathcal C}_3=2$}
\psfrag{c4}{\tiny ${\mathcal C}_4=\bar 2$}
 \begin{minipage}{0.18\textwidth}
   \centering
  \includegraphics[width=170pt]{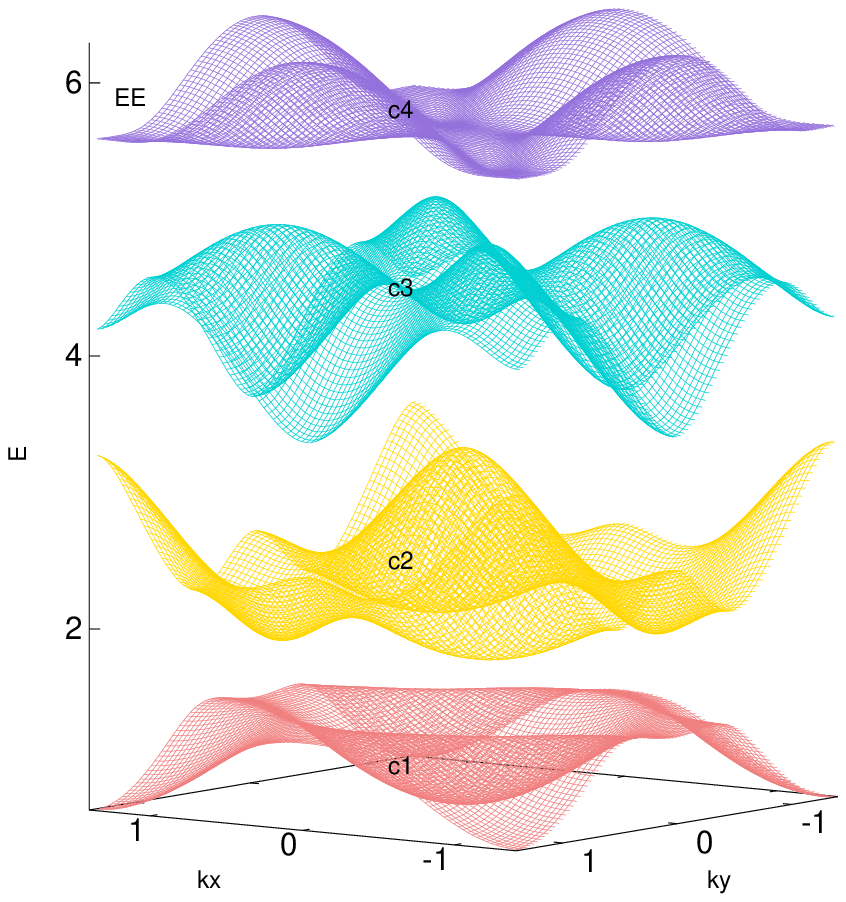}
   \end{minipage}\hfill
   \begin{minipage}{0.28\textwidth}
\psfrag{E}{}
\psfrag{EE}{(f)}
    \psfrag{c1}{\tiny ${\mathcal C}_1=0$}
\psfrag{c2}{\tiny ${\mathcal C}_2=2$}
\psfrag{c3}{\tiny ${\mathcal C}_3=\bar 2$}
\psfrag{c4}{\tiny ${\mathcal C}_4=0$}
    \centering
  \includegraphics[width=170pt]{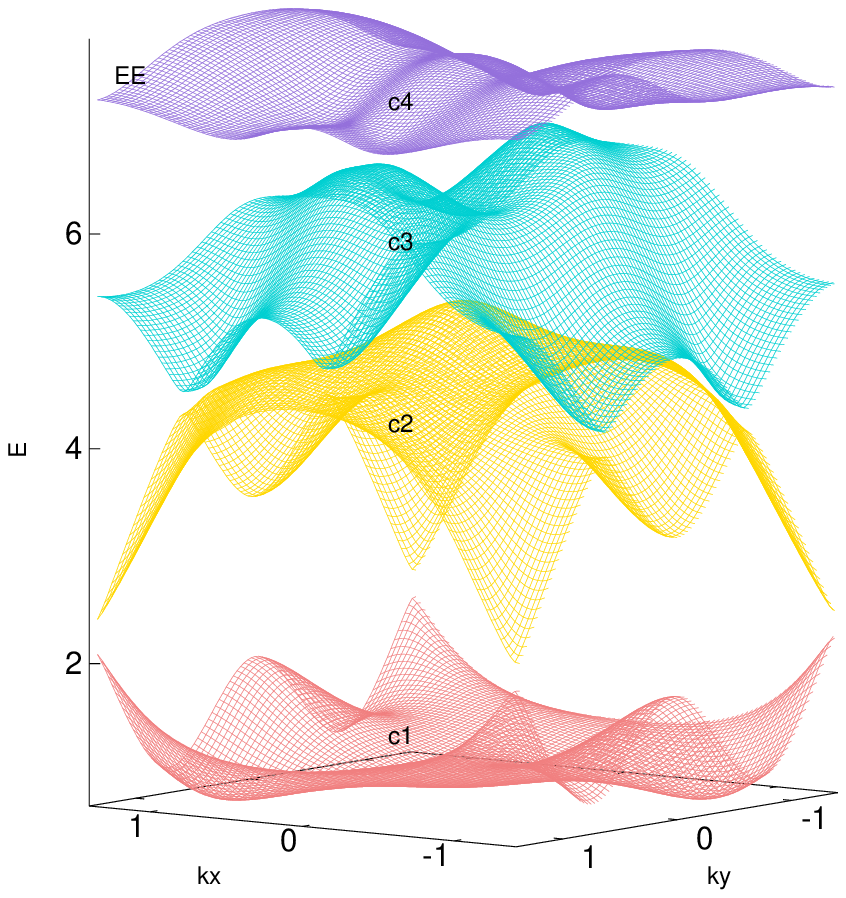}
   \end{minipage}\hfill
  \caption{The magnon bands. 
a) $J=-1$, $D_{\rm m}=0.5$ for ${\mathcal C}=(11\bar 1\bar 1)$, 
(b) $J=-1$, $K'=-0.5$, $D_{\rm m}=0.5$ for ${\mathcal C}=(11\bar 20)$, 
(c) $J=-1$, $K=0.5$, $D_{\rm m}=0.5$ for ${\mathcal C}=(1\bar 11\bar 1)$, 
(d) $J=-1$, $J^\prime=-0.5$, $K'=-1$, $D_{\rm m}=1$ for ${\mathcal C}=(002\bar 2)$, 
(e)  $J^\prime=-0.5$, $K'=-0.5$, $D_{\rm m}=1$ for ${\mathcal C}=(\bar 222\bar 2)$, and 
(f) $J=-1$, $K=0.5$, $K'=-1$, $D_{\rm m}=1$ for ${\mathcal C}=(02\bar 20)$, 
with $h=1$.
Parameters of non-zero values are indicated here.}
   \label{energy3d}
  \end{figure}    
   
 The magnon dispersion relations based on the 
exact FM ground state are obtained by converting the spin 
operators in terms of bosonic creation ($b^\dagger$) 
and annihilation ($b$) operators via the Holstein-Primakoff transformation, 
\begin{equation}
 \begin{aligned}
  S_j^z=S-b_j^\dagger b_j,\;\;
  S_j^+\simeq\sqrt{2S}\, b_j,\;\;
  S_j^-\simeq\sqrt{2S}\, b_j^\dagger. 
 \end{aligned}
\end{equation}
Hamiltonian (Eq \ref{ham}) has been expressed  
in the momentum space by Fourier-transforming the 
operators in that space, 
$b_j=\frac{1}{\sqrt N}\,\sum_{\boldsymbol{k}}b_{\boldsymbol{k}}\,e^{i\,\boldsymbol{k}\cdot 
\boldsymbol{R}_j}$, where $N$ is the total number of primitive cells. Thus, 
the Hamiltonian with respect to the ground state energy, 
$E_{\rm G}=2NS^2[3(J+J')+2(K_z+K_z')]-4NS$, is
\begin{equation}
 \begin{aligned}
{\mathcal H}=\!\frac{S}{2}\sum\limits_{\bold{k}}\Psi^\dagger_{\bold{k}}
{\mathcal H}_\bold{k}\Psi_{\bold{k}}, 
 \end{aligned}
\label{hhexa}
 \end{equation}
 where,  
$\Psi^\dagger_{\bold{k}}=[b^\dagger_{A,\bold{k}},b^\dagger_{B,\bold{k}}b^\dagger_{C,\bold{k}},b^\dagger_{D,\bold{k}},b_{A,-\bold{k}},b_{B,-\bold{k}},b_{C,-\bold{k}},b_{D,-\bold{k}}]$. 
Now, $b^\dagger_A$, $b^\dagger_B$, $b^\dagger_C$ and $b^\dagger_D$ are the bosonic creation operators 
on the sublattices $A$, $B$, $C$ and $D$, respectively. Terms containing product of four bosonic 
operators have been neglected since they invoke inter-magnon interactions. 
${\mathcal H}_\bold{k}$ is a $8\!\times \!8$ matrix, which  can be expressed 
in terms of two different $4\! \times \!4$ matrices, $X_\bold{k}$ and $Y_\bold{k}$ as 
 \begin{equation}
 \begin{aligned}
{\mathcal H}_\bold{k}=
 \left[ 
 { \begin{array}{cc}
  X_\bold{k}  &  Y_\bold{k}  \\
  Y^\dagger_\bold{k} &   X^T_{-\bold{k}}  \\
  \end{array}}
\right]\!,\,
\label{Hk}
 \end{aligned}
 \end{equation}
  where, both $X_\bold{k}$ and $Y_\bold{k}$ are Hermitian. 
  \begin{equation}
 \begin{aligned} 
X_\bold{k}=
 \left[ 
 { \begin{array}{cccc}
  a_0  &   a_{1,\bold{k}} &   a_{2,\bold{k}} &   a_{3,\bold{k}}  \\
  a^*_{1,\bold{k}} &   a_0  &   a_{4,\bold{k}}  &   a_{5,\bold{k}}  \\
  a^*_{2,\bold{k}} &   a^*_{4,\bold{k}}  &   a_0  &   a_{6,\bold{k}}  \\
   a^*_{3,\bold{k}} &   a^*_{5,\bold{k}}  &   a^*_{6,\bold{k}}  &   a_0  \\
  \end{array}}
\right]\!, \,\\
Y_\bold{k}=
 \left[ 
{ \begin{array}{cccc}
  0  &   b_{1,\bold{k}} &   b_{2,\bold{k}} &   b_{3,\bold{k}}  \\
  b_{1,-\bold{k}} &   0  &   b_{4,\bold{k}}  &   b_{5,\bold{k}}  \\
  b_{2,-\bold{k}} &   b_{4,-\bold{k}}  &   0  &   b_{6,\bold{k}}  \\
   b_{3,-\bold{k}} &   b_{5,-\bold{k}}  &   b_{6,-\bold{k}}  &   0  \\
  \end{array}}
\right]\!.
\end{aligned}
 \end{equation}
Considering the spin polarization along the $+z$ direction 
components of $X_\bold{k}$ and $Y_\bold{k}$ are obtained below.
\begin{equation}
  \begin{aligned}
  a_0=&-3J-2K_z-3J^\prime-2K^\prime_z+h/S, \\
  a_{1,\bold{k}}=&\,(J+K_y)+(J^\prime+K^\prime_y) e^{i\bold{k}\cdot\boldsymbol{\delta}_1},\\
a_{2,\bold{k}}=&\,(J+K_x) e^{i\bold{k}\cdot\boldsymbol{\delta}_1}+(J^\prime+K^\prime_x),\\
a_{3,\bold{k}}=&\,J+J^\prime e^{-i\bold{k}\cdot\boldsymbol{\delta}_2}-iD_{\rm m} e^{-i\bold{k}\cdot\boldsymbol{\delta}_2},\\
a_{4,\bold{k}}=&\,J+J^\prime e^{-i\bold{k}\cdot\boldsymbol{\delta}_2}+iD_{\rm m} e^{-i\bold{k}\cdot\boldsymbol{\delta}_2},\\
a_{5,\bold{k}}=&\,(J+K_x) e^{-i\bold{k}\cdot\boldsymbol{\delta}_2}+(J^\prime+K^\prime_x),\\
 a_{6,\bold{k}}=&\,(J+K_y)+(J^\prime+K^\prime_y) e^{-i\bold{k}\cdot\boldsymbol{\delta}_1},\\
   b_{1,\bold{k}}=&-K_y-K^\prime_y e^{i\bold{k}\cdot\boldsymbol{\delta}_1},\\
b_{2,\bold{k}}=&\,K_x e^{i\bold{k}\cdot\boldsymbol{\delta}_1}+K^\prime_x,\\
b_{3,\bold{k}}=&\,b_{4,\bold{k}}=0,\\
b_{5,\bold{k}}=& \,K_x e^{-i\bold{k}\cdot\boldsymbol{\delta}_2}+K^\prime_x,\\
 b_{6,\bold{k}}=&-K_y-K^\prime_y e^{-i\bold{k}\cdot\boldsymbol{\delta}_1},\;{\rm with}, \\
\boldsymbol{\delta}_1=&\,a \,\sqrt{5}\;\hat{i}, \;
  \boldsymbol{\delta}_2=a\,\sqrt{5}\;\hat{j}. \nonumber
  \end{aligned}
\end{equation} 
Here, $a$ is the length of NN bond which is assumed to be unity. 
Following the Bogoliubov diagonalization method applicable for 
the bosonic operators, the non-Hermitian matrix, $I_{\rm B} {\mathcal H}_\bold{k}$ 
has been diagonalized, instead of ${\mathcal H}_\bold{k}$, in order to 
obtain the eigenenergies and eigenmodes where
$I_{\rm B}=$ diag[$1,1,1,1,-1,-1,-1,-1$]. 
Four positive eigenenergies of $I_{\rm B} {\mathcal H}_\bold{k}$ 
are thus treated as the magnon excitation energies of the system. 
Accuracy of the results increases with the value of $S$. 
Topological phases have been obtained  
in the regime where the real eigenenergies are available. 
Eigenenergies thus constitute the four-band magnon dispersion relations. 
Band structure varies with the value of magnetic field in such 
a fashion that no alteration of the topological phases is found. 

  \begin{figure}[h]
  \psfrag{E}{ $E$}
\psfrag{X}{X}
\psfrag{G}{  $\Gamma$}
\psfrag{K}{ $\bold{k}$}
\psfrag{M}{ M}   
   \psfrag{AA}{\scriptsize(a)}
    \begin{minipage}{0.21\textwidth}
   \psfrag{c1}{\tiny ${\mathcal C}_1=1$}
\psfrag{c2}{\tiny ${\mathcal C}_2=1$}
\psfrag{c3}{\tiny ${\mathcal C}_3=\bar 1$}
\psfrag{c4}{\tiny ${\mathcal C}_4=\bar 1$}
    \centering
  \includegraphics[width=125pt]{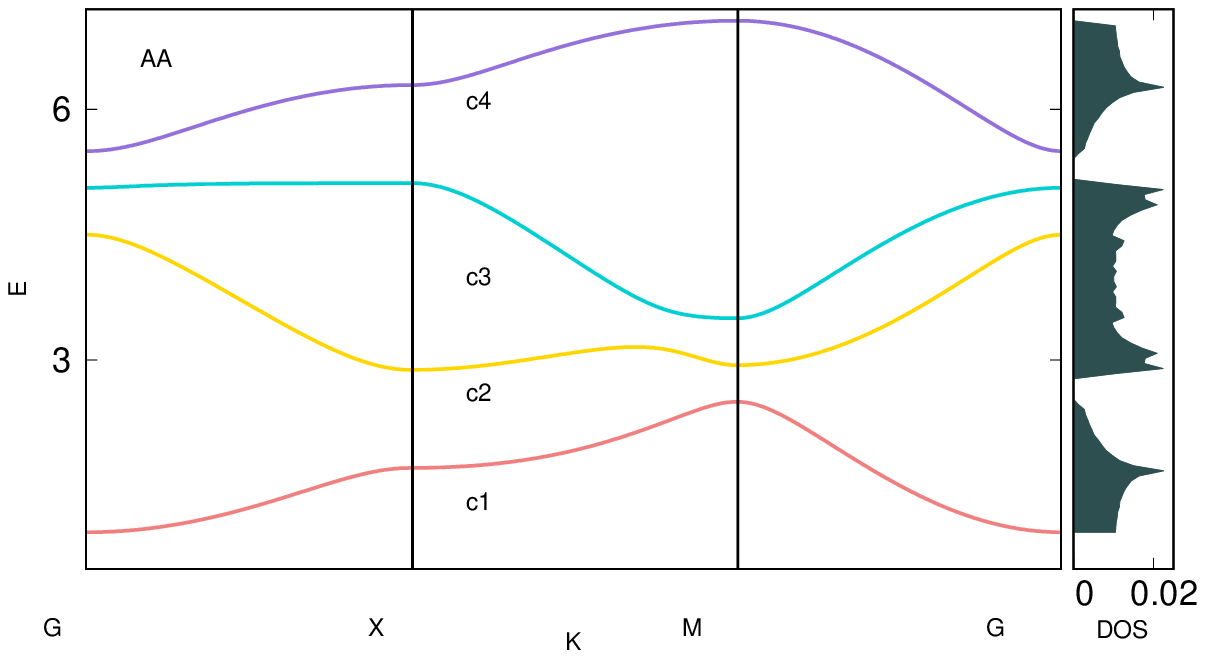}
 \end{minipage}\hfill   
  \begin{minipage}{0.25\textwidth}
   \psfrag{BB}{\scriptsize(b)}
   \psfrag{c1}{\tiny ${\mathcal C}_1=1$}
\psfrag{c2}{\tiny ${\mathcal C}_2=1$}
\psfrag{c3}{\tiny ${\mathcal C}_3=\bar 2$}
\psfrag{c4}{\tiny ${\mathcal C}_4=0$}
 \psfrag{E}{ }
     \centering
  \includegraphics[width=125pt]{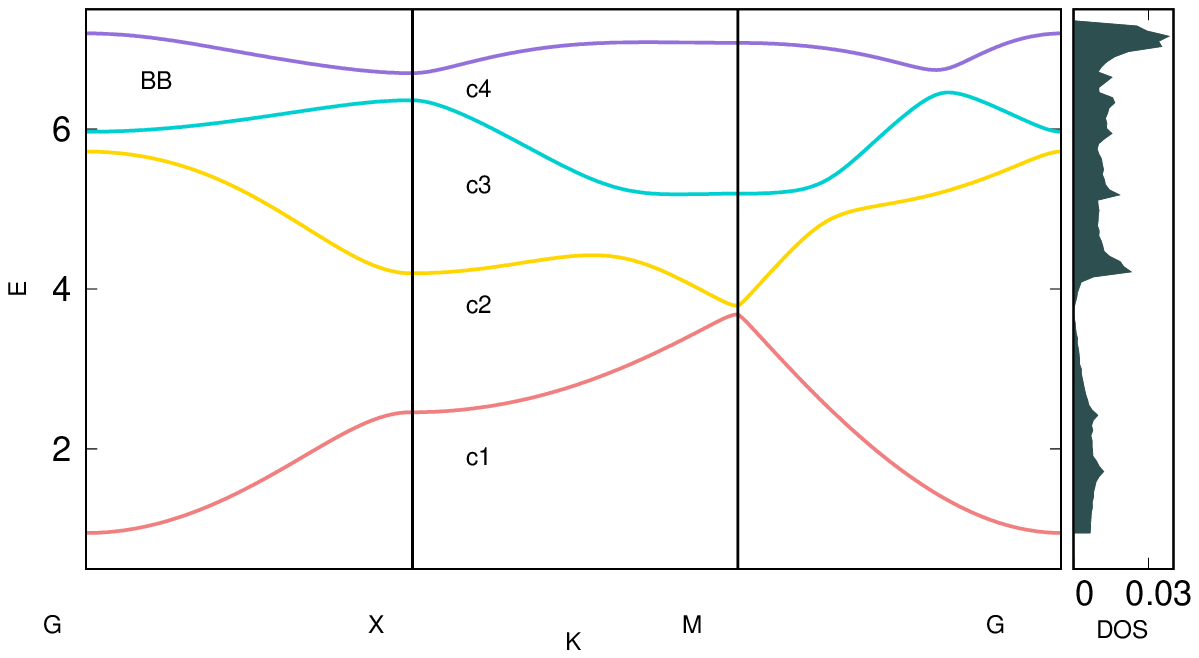}
  \end{minipage}\hfill   
  
   \psfrag{CC}{\scriptsize(c)}
    \psfrag{E}{E }
     \begin{minipage}{0.21\textwidth}
  \psfrag{c1}{\tiny ${\mathcal C}_1=1$}
\psfrag{c2}{\tiny ${\mathcal C}_2=\bar 1$}
\psfrag{c3}{\tiny ${\mathcal C}_3=1$}
\psfrag{c4}{\tiny ${\mathcal C}_4=\bar 1$}
    \centering
  \includegraphics[width=125pt]{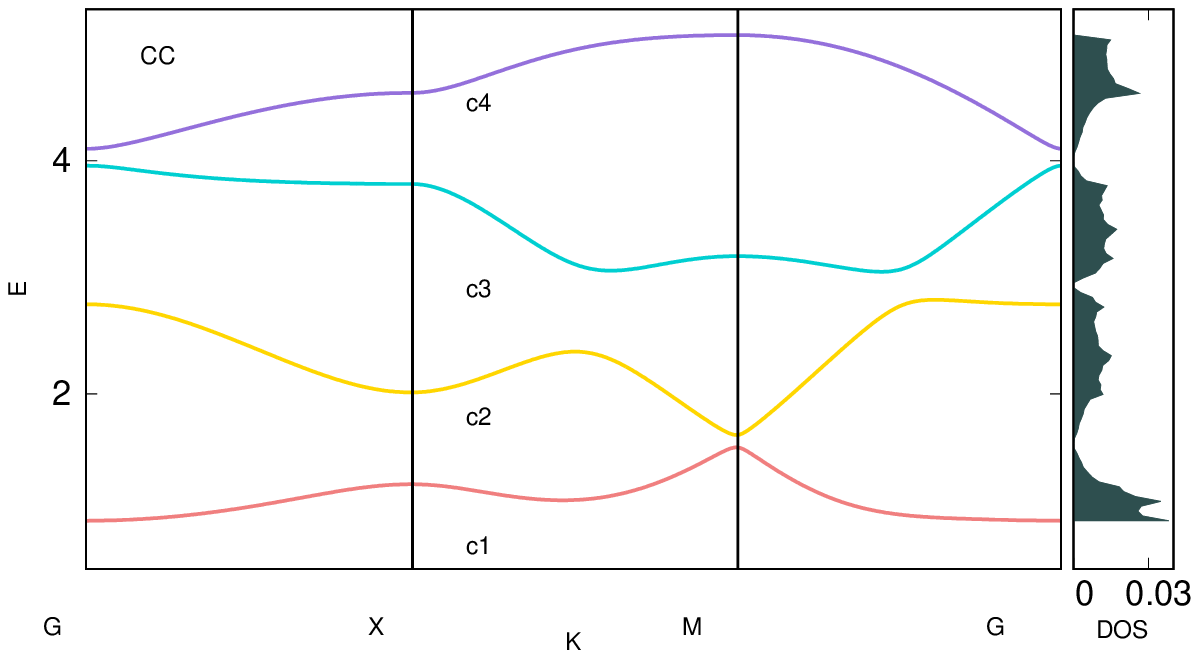}
  \end{minipage}\hfill   
  \begin{minipage}{0.25\textwidth}
 \psfrag{DD}{\scriptsize(d)}
  \psfrag{E}{ }
  \psfrag{c1}{\tiny ${\mathcal C}_1=0$}
\psfrag{c2}{\tiny ${\mathcal C}_2=0$}
\psfrag{c3}{\tiny ${\mathcal C}_3=2$}
\psfrag{c4}{\tiny ${\mathcal C}_4=\bar 2$}
     \centering
  \includegraphics[width=125pt]{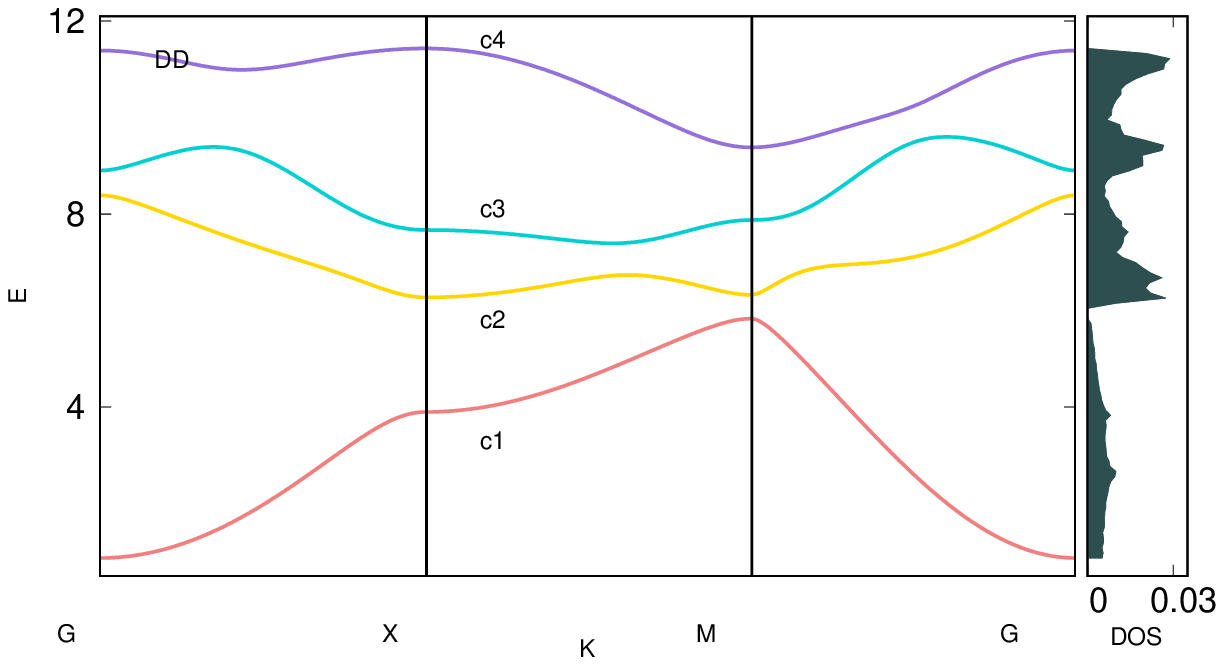}
  \end{minipage}\hfill   
  
    \psfrag{EE}{(e)} 
     \psfrag{E}{E }
  \begin{minipage}{0.21\textwidth}
   \psfrag{c1}{\tiny $C_1=\bar 2$}
\psfrag{c2}{\tiny $C_2=2$}
\psfrag{c3}{\tiny $C_3=2$}
\psfrag{c4}{\tiny $C_4=\bar 2$}
 \includegraphics[width=125pt]{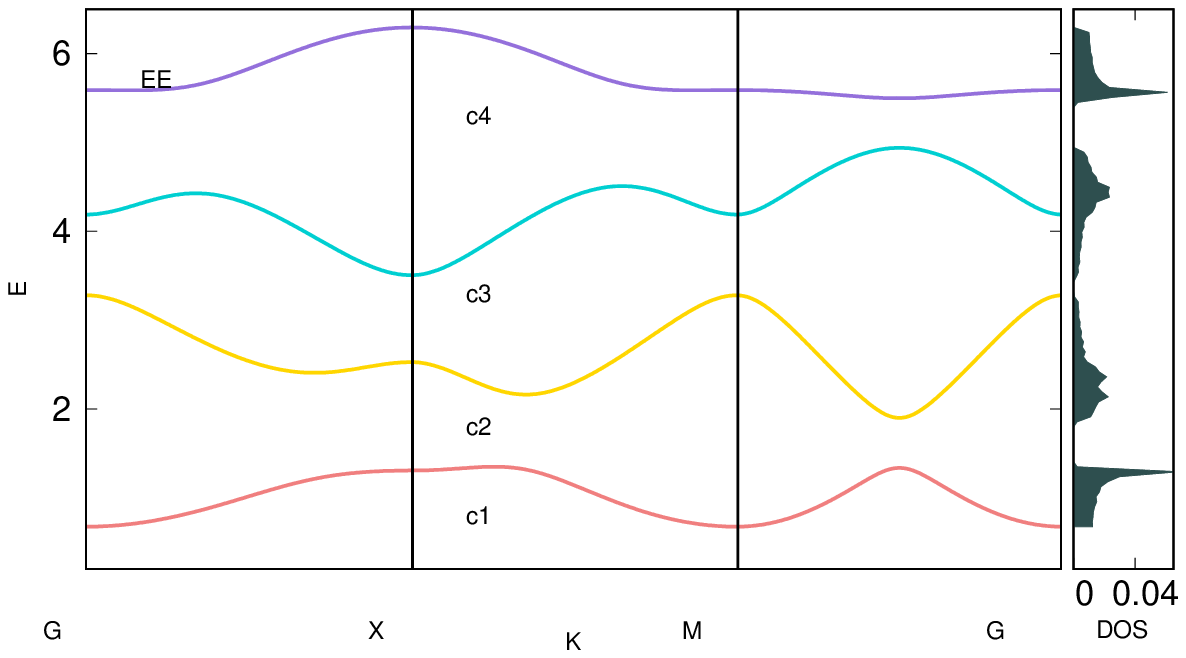}
 \end{minipage}\hfill 
 \begin{minipage}{0.25\textwidth}
 \psfrag{EE}{\scriptsize(f)}
  \psfrag{E}{ }
  \psfrag{c1}{\tiny $C_1=0$}
\psfrag{c2}{\tiny $C_2=2$}
\psfrag{c3}{\tiny $C_3=\bar 2$}
\psfrag{c4}{\tiny $C_4=0$}
     \centering
  \includegraphics[width=125pt]{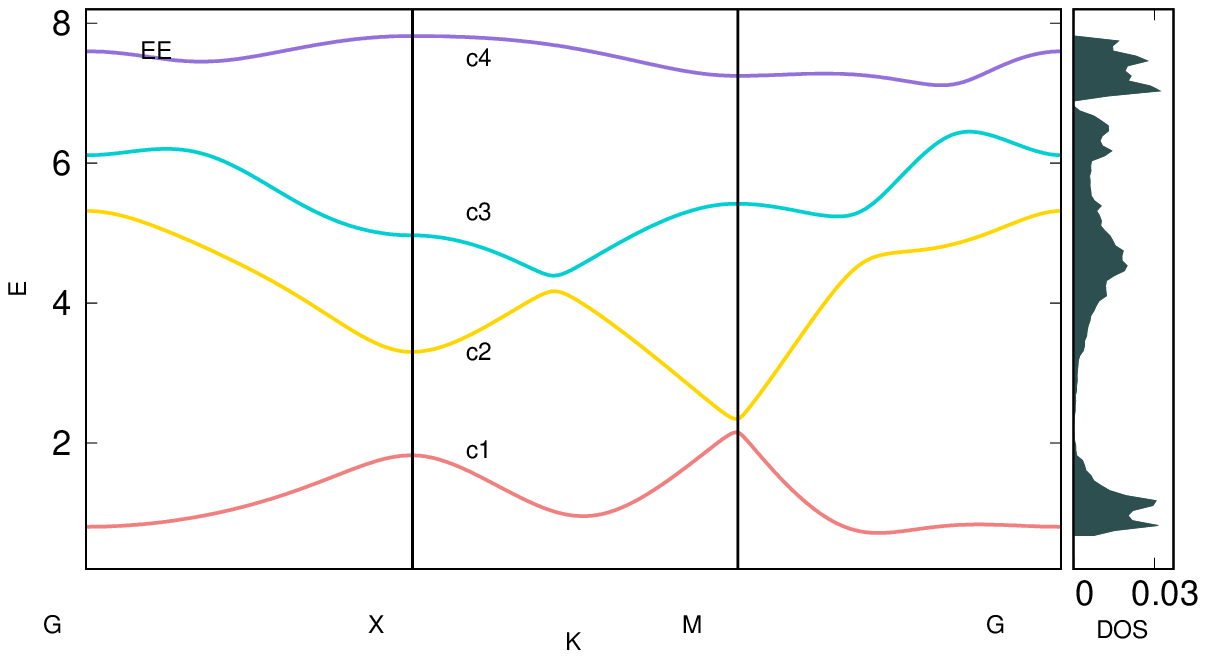}
  \end{minipage}\hfill   
   
  \caption{Dispersion relation along the high-symmetry points of Brillouin zone. 
The side panel shows the DOS. Values of the parameters are the 
same as Fig \ref{energy3d} for the respective plots.}
     \label{energybz}
  \end{figure}            
\section{Chern number, edge states, thermal Hall conductance AND THE
TOPOLOGICAL PHASES}
\label{CETHC}
In order to characterize topological phases of this  
four-bands system, the Chern number for each distinct magnon band is obtained, when 
there remains a definite gap between adjacent bands. 
Chern number for
the $i$-th band, ${\mathcal C}_i$, is obtained by integrating 
the Berry curvature of that band, ${\mathcal F}_i(\bold{k})$ 
over the BZ.
\begin{equation}
  \begin{aligned}
   {\mathcal C}_i=\frac{1}{2\pi }\iint_{\rm BZ} {\mathcal F}_i(\bold{k})\,d^2\bold{k},
  \end{aligned}
\end{equation}
where, ${\mathcal F}_i(\bold{k})$ is expressed 
in terms of the corresponding Berry connection, 
${\mathcal A}^i_{\mu}(\bold{k}) = \braket{\psi_i(\bold{k})|\partial_{k_{\mu}}|\psi_i(\bold{k})}$ as 
${\mathcal F}_i (\bold{k})= {\partial_{ k_x}} {\mathcal A}^i_{y}(\bold{k})-
{\partial_{ k_y}} {\mathcal A}^i_{x}(\bold{k})$, and 
$|\psi^i(\bold{k})\rangle$ is the eigenvector of the $i$-th magnon band. 
${\mathcal C}_i$ has been evaluated numerically \cite{Fukui}.       
   
According to the `bulk-edge-correspondence' rule the 
presence of nonzero value of ${\mathcal C}$ implies the existence of edge states. 
Edge state corresponds to the surface property of the system. To calculate
the bulk-edge energy spectrum a pair of edges parallel to the 
$x$-axis is created here by breaking the 
PBC along the $y$-axis. As a result,  
a strip of CaVO lattice is constructed which has $N$ primitive cells 
along the $\hat{y}$ and infinitely long 
towards the $\hat{x}$. The corresponding lattice structure is shown in Fig \ref{lattice}(b). Fourier
transform of the bosonic operators is taken only along the $x$ direction and  
4$N\times 4N$ Hamiltonian has been obtained. 

THC, $\kappa_{xy}$, of the system can be expressed in 
terms of ${\mathcal F}(\bold{k})$ for the system as \cite{Murakami1,Murakami2}, 
   \begin{equation}
  \begin{aligned}
   \kappa_{xy}(T)=-\frac{k^2_B T}{4\pi^2\hbar}\,\sum\limits_{i}\, \iint_{\rm BZ} 
c(\rho_{i}(\bold{k}))\,{\mathcal F}_{i}(\bold{k})\,\,d^2\bold{k}. 
  \end{aligned}
\end{equation}
Here $T$ is the temperature, $k_B$ is the Boltzmann constant 
and $\hbar$ is the reduced 
Planck's constant. 
$c(x)=(1+x)\left(\ln{\frac{1+x}{x}}\right)^2-\left(\ln x\right)^2 -2{\rm Li}_2(-x)$, where
${\rm Li}_2(z)=-\int_{0}^{z} du \frac{\ln{(1-u)}}{u}$ 
and $ \rho_{i}(\bold{k})$ is the Bose-Einstein distribution, {\em i.e.},
$\rho_{i}(\bold{k})=1/(e^{E^i_{\bold{k}}/k_{B}T} -1)$. 
Like all thermodynamic quantities, $\kappa_{xy}(T)$ also gets saturated at high temperatures. 
As $\kappa_{xy}(T)$ directly depends on the Berry curvature so it behaves differently in different
topological phases. As a result, $\kappa_{xy}$ 
suffers sudden change in its value at the phase transition points. 

In this study, isotropic Heisenberg interaction on the 
NN and NNN bonds is assumed, while 
both isotropic and anisotropic Kitaev couplings are taken into account. 
Anisotropic XXZ Heisenberg interaction on the 
NN and NNN bonds is not considered here because of the following reason. 
Total Hamiltonian containing isotropic Heisenberg and DMI terms 
can be mapped on to the anisotropic XXZ Heisenberg Hamiltonian via a 
canonical transformation of the spin operators, which is valid  
for any values of $S$ \cite{Wreszinski}.
It is true even for arbitrary directions of 
$\boldsymbol{D}_{\rm m}$ over different bonds as long as 
they are parallel or antiparallel to each other \cite{Maleyev}.
Which means that effect of DMI term on the Heisenberg model 
can be studied in terms of a suitable XXZ Heisenberg Hamiltonian 
where the value of anisotropic parameter depends on the 
values of exchange and DMI strengths. 
So, in other words, inclusion of anisotropic Heisenberg interaction
could not lead to the emergence of new topological phases anymore. 

For example, the same set of two distinct topological phases with 
 ${\mathcal C}$=($10\bar{1}$) and ($\bar 101$),
appear in two previously studied models 
where isotropic and anisotropic XXZ Heisenberg Hamiltonians 
are formulated on kagom\'e lattice in the presence of DMI \cite{Li,Sen}. 
A closer scrutiny on
those two models 
reveals that only the diagonal terms of those 
Hamiltonian matrices ($H_{ij}$) are different, where the
values of ${\mathcal C}$s are insensitive to them. Off-diagonal matrix elements satisfy the 
relation, $H_{ij}(-D_{\rm m})=H_{ij}^*(D_{\rm m})$, 
which on the other hand corresponds to the   
band inversion about $D_{\rm m}=0$, in this particular case \cite{Li}. 
And as a result, ${\mathcal C}$s of the two 
topological phases exhibit mirror symmetry around the middle band. 

Here, the system hosts seventeen distinct topological phases in total. Six are found 
for the isotropic Kitaev coupling but twelve for the anisotropic case. All of them are 
described in the following two subsections. 
One phase is found common in both the cases. 
Every topological phase 
is described by band structure, dispersion relation along the high-symmetric points of 
first Brillouin zone, density of states (DOS) and thermal Hall conductance.
Value of ${\mathcal C}$ for each distinct band has been evaluated in 
association with the bulk-edge energy spectrum  
for the Hamiltonian formulated on the strip of CaVO 
lattice of finite length along $y$-axis.
\begin{figure}[h]
 \psfrag{E}{$E$}
 \psfrag{qx}{\scriptsize $k_x$}
 \psfrag{w}{ \scriptsize$|\psi|^2$}
 \psfrag{AA}{\scriptsize(a)}
 \psfrag{BB}{\scriptsize(b)}
\psfrag{CC}{\scriptsize(c)}
 \psfrag{DD}{\scriptsize(d)}
 \psfrag{EE}{\scriptsize(e)}
 \psfrag{site}{\scriptsize site}
 \psfrag{ub}{\tiny upper edge}
 \psfrag{lb}{\tiny lower edge}
 \begin{minipage}{0.21\textwidth}
   \includegraphics[width=130pt]{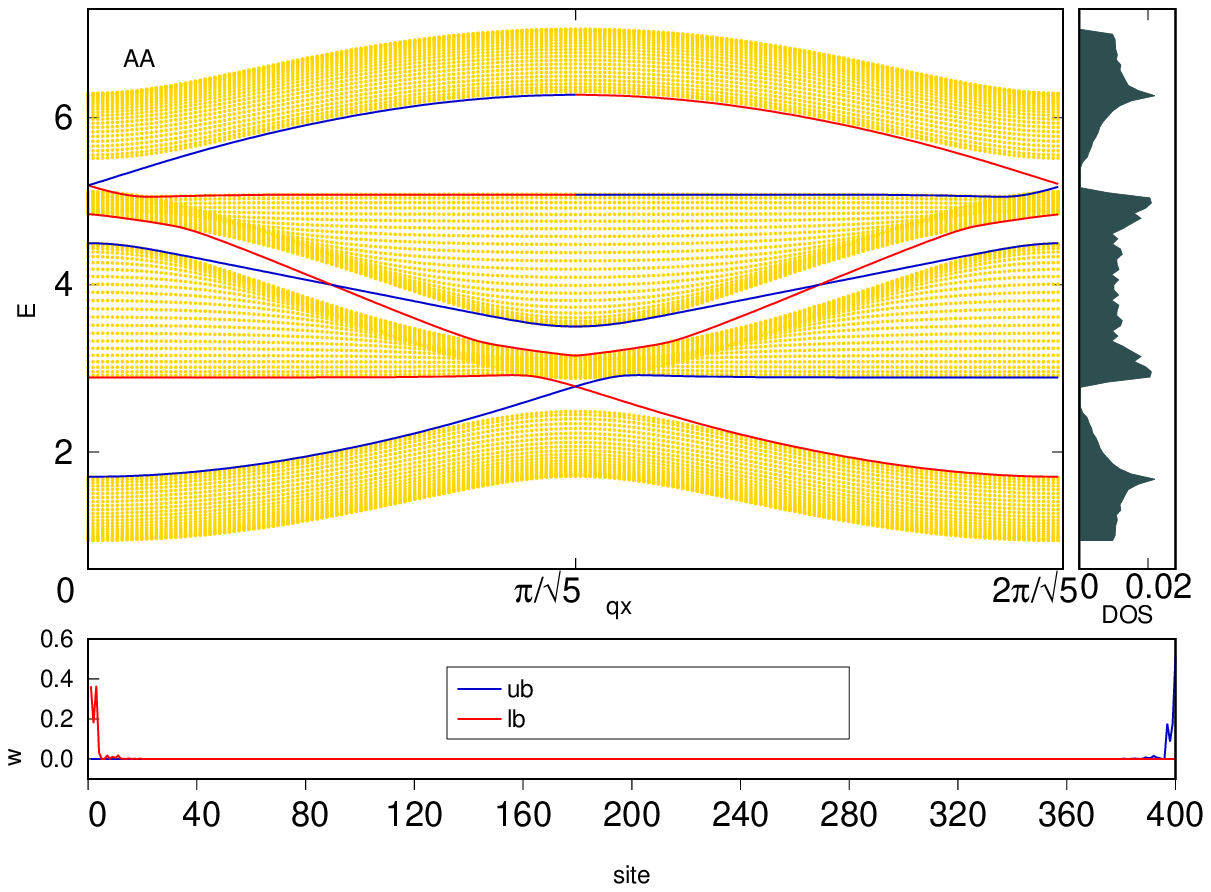}
   \end{minipage}\hfill
   \begin{minipage}{0.24\textwidth}
     \psfrag{E}{}
   \psfrag{w}{}
   \includegraphics[width=130pt]{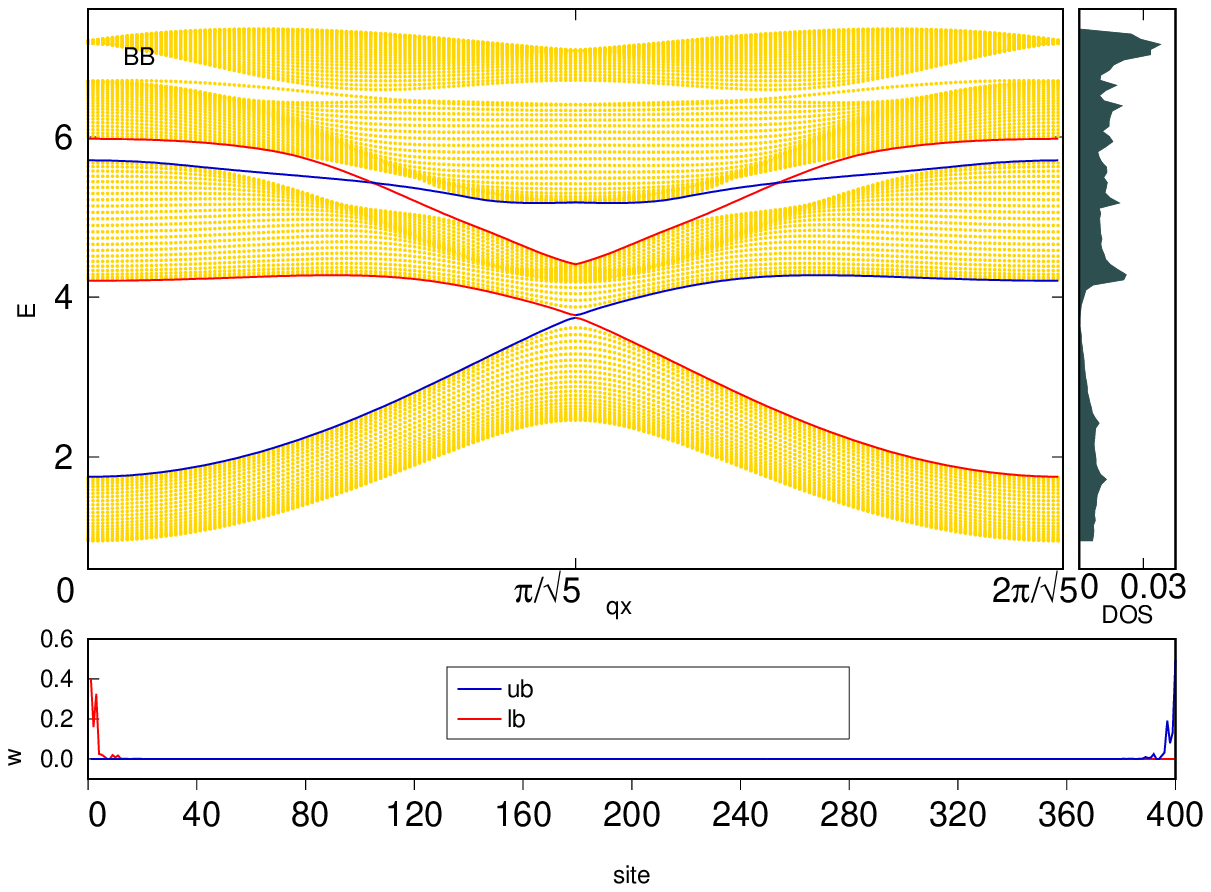}
   \end{minipage}\hfill
   \vskip 0.1cm
     \psfrag{E}{$E$}
   \psfrag{w}{\scriptsize $|\psi|^2$}
   \begin{minipage}{0.21\textwidth}
   \includegraphics[width=130pt]{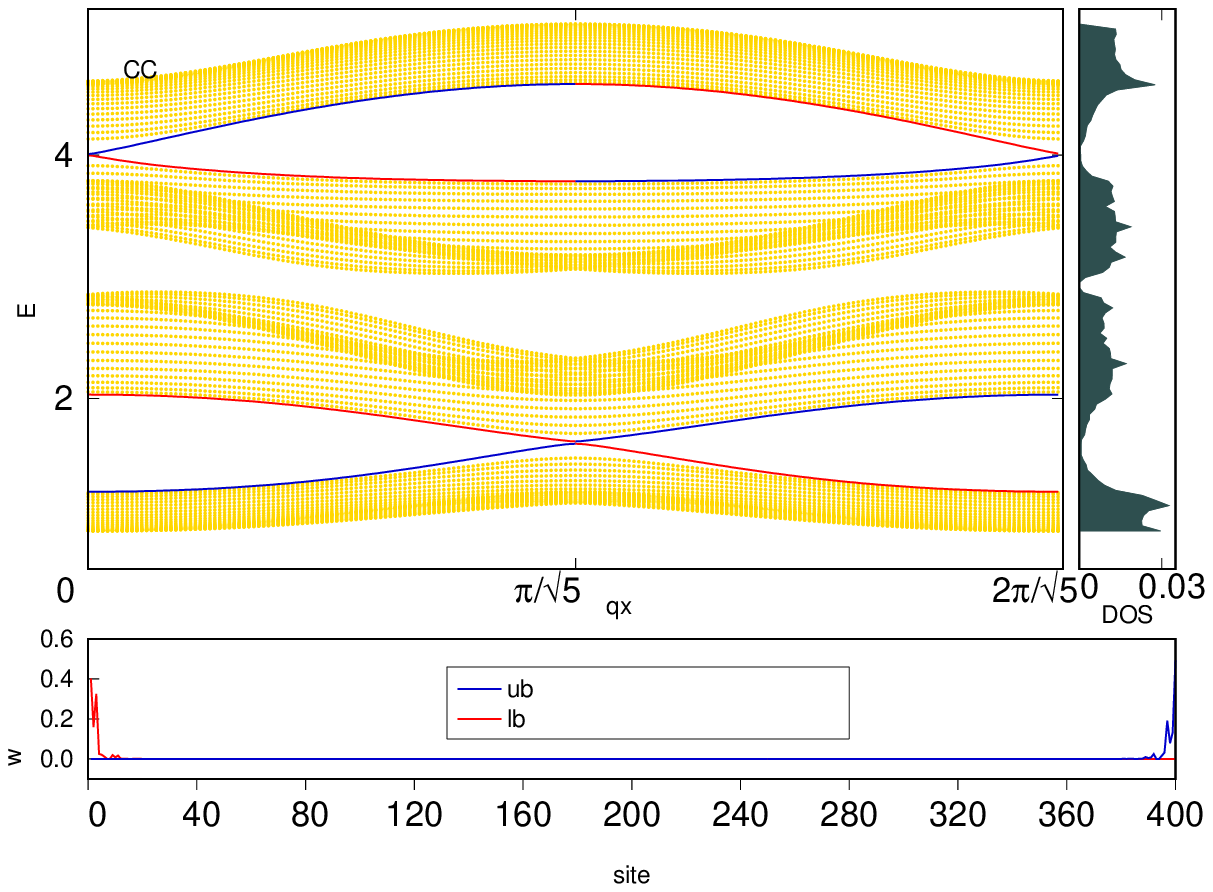}
    \end{minipage}\hfill
   \begin{minipage}{0.24\textwidth}
    \psfrag{E}{}
   \psfrag{w}{}
   \includegraphics[width=130pt]{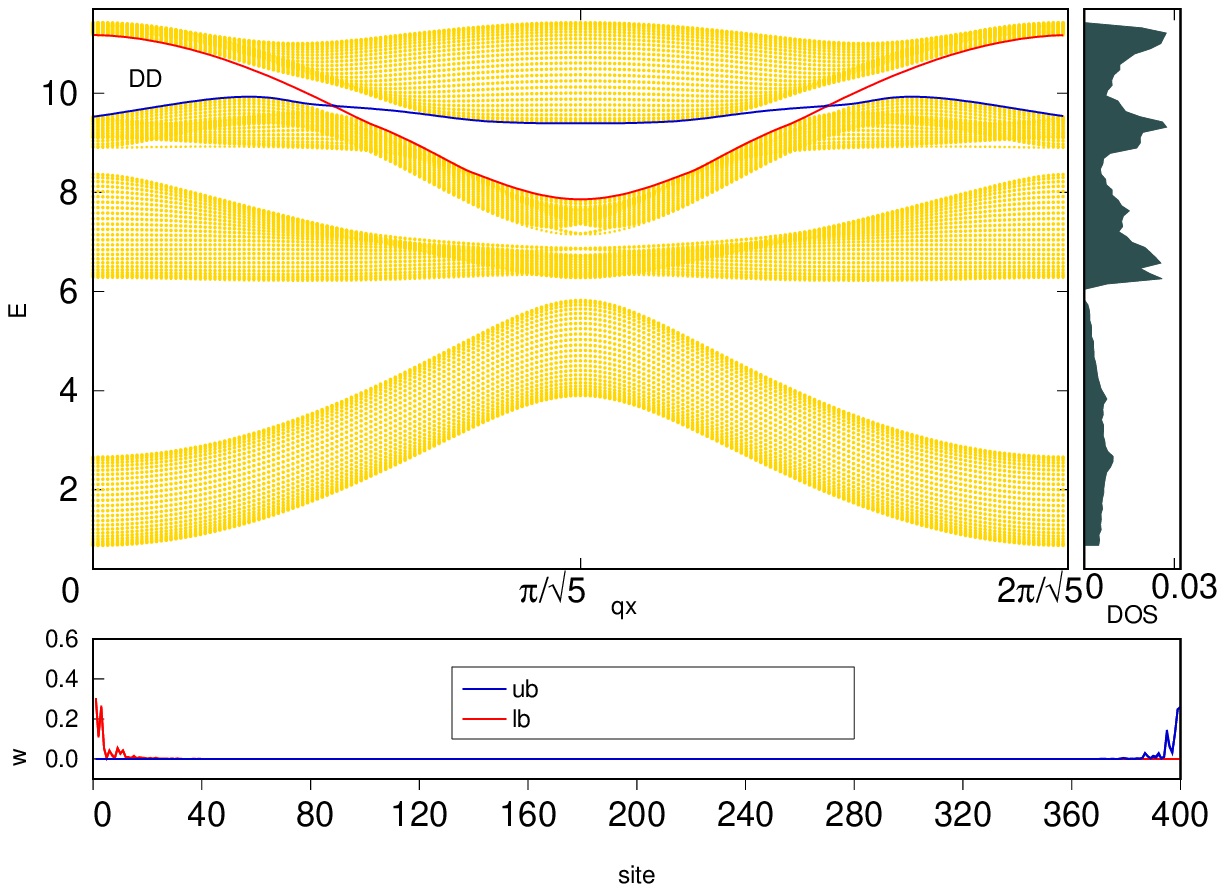}
   \end{minipage}\hfill
   \vskip 0.1cm
     \psfrag{E}{$E$ }
   \psfrag{w}{\scriptsize $|\psi|^2$}
    \begin{minipage}{0.21\textwidth}
   \includegraphics[width=130pt]{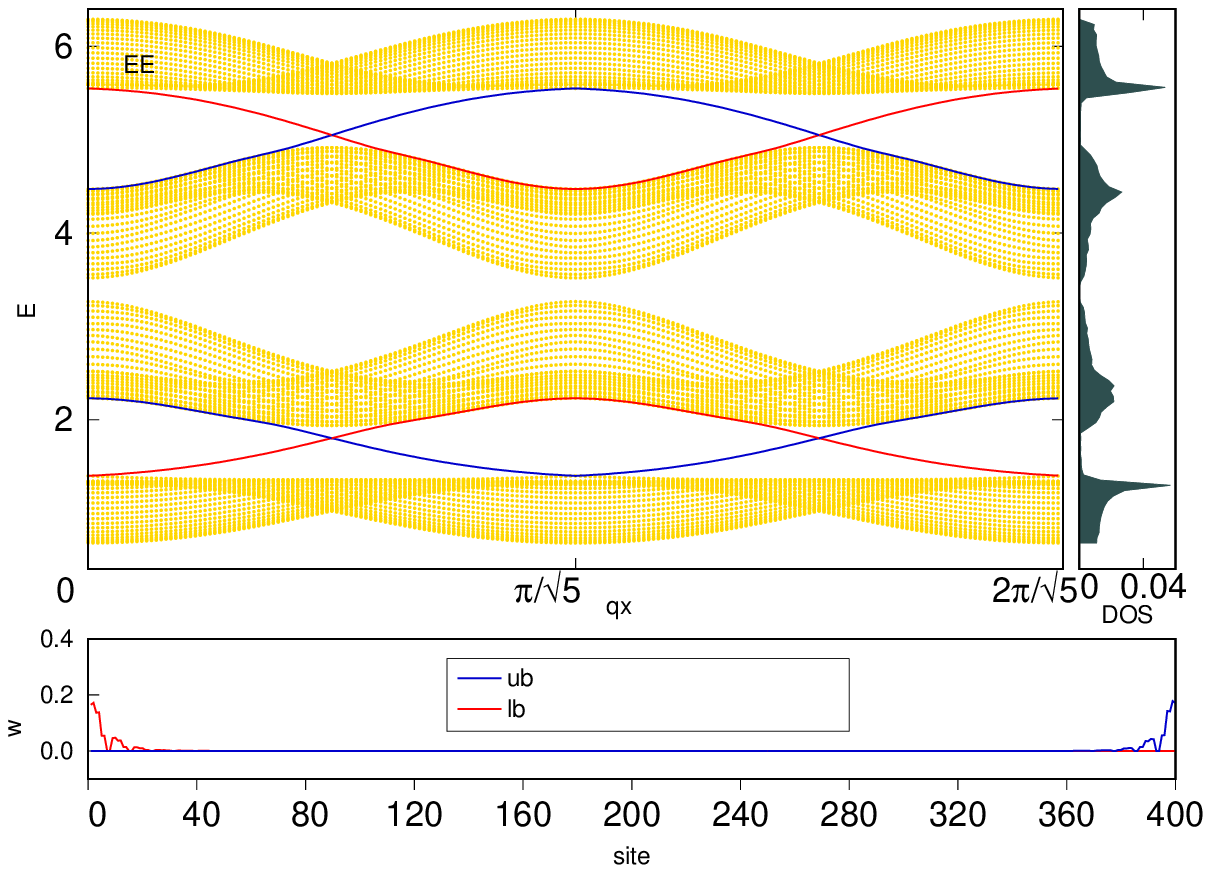}
   \end{minipage}\hfill
   \begin{minipage}{0.24\textwidth}
    \psfrag{E}{}
   \psfrag{w}{}
   \psfrag{DD}{\scriptsize(f)}
   \includegraphics[width=130pt]{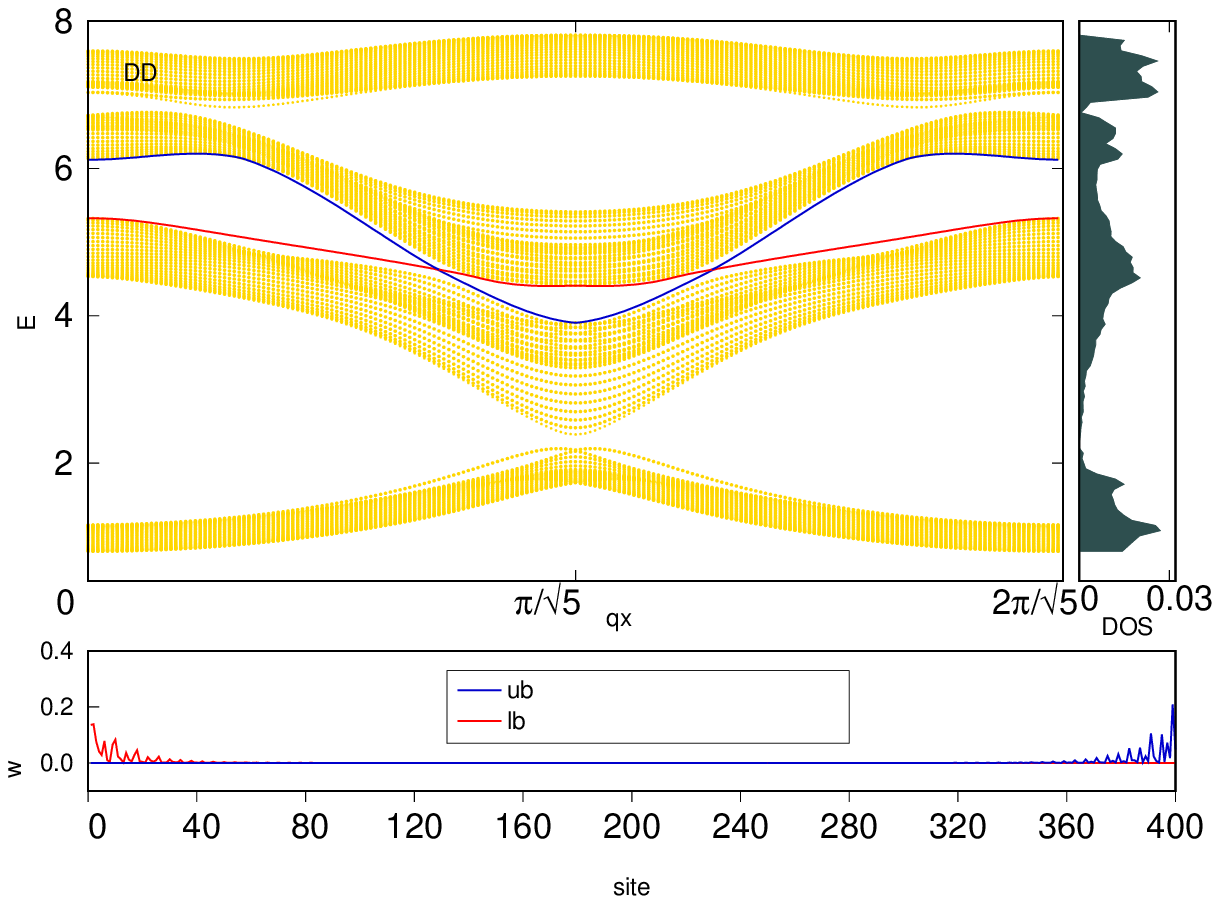}
   \end{minipage}\hfill
  \caption{Magnon dispersions of bulk-edge states in the one-dimensional BZ. 
Upper and lower edge modes are drawn in blue and red lines, respectively, while 
bulk modes are in golden points. 
The side panel shows the DOS.
 The lower panel indicates variation of probability density of both edge
modes with respect to site number for a fixed $k_x$. Values of the parameters are:  
a) $J=-1$, $D_{\rm m}=0.5$ for ${\mathcal C}=(11\bar 1\bar 1)$, 
(b) $J=-1$, $K'=-0.5$, $D_{\rm m}=0.5$ for ${\mathcal C}=(11\bar 20)$, 
(c) $J=-1$, $K=0.5$, $D_{\rm m}=0.5$ for ${\mathcal C}=(1\bar 11\bar 1)$, 
(d) $J=-1$, $J^\prime=-0.5$, $K'=-1$, $D_{\rm m}=1$ for ${\mathcal C}=(002\bar 2)$, 
(e)  $J^\prime=-0.5$, $K'=-0.5$, $D_{\rm m}=1$ for ${\mathcal C}=(\bar 222\bar 2)$, and 
(f) $J=-1$, $K=0.5$, $K'=-1$, $D_{\rm m}=1$ for ${\mathcal C}=(02\bar 20)$, 
with $h=1$. Parameters with only nonzero values are mentioned here.}
 \label{edge}
   \end{figure}       
\subsection{Isotropic Kitaev coupling}
In this case, the same value of Kitaev interaction along three different links of the 
lattice is considered, which means, $K_\gamma=K$ and $K^\prime_\gamma=K^\prime$. 
Similar model on the two-band honeycomb lattice 
exhibits multiple TMI phase with higher values of ${\mathcal C}$s,  
when third neighbor interactions are invoked \cite{Moumita}. 
In this study, isotropic Kitaev model on the CaVO 
lattice is found to exhibit six topological phases. 
Bulk energy dispersion with specific values of ${\mathcal C}$ have been plotted 
in Fig \ref{energy3d} for different TMI phases.   
Dispersion relations along the 
high-symmetry points of BZ, in addition to the DOS are 
shown in Fig \ref{energybz}. DOS 
confirms the existence of band gap in every case. 
 Dispersions of bulk-edge states in the one-dimensional BZ 
 have been shown in Fig \ref{edge}. 
Edge state dispersion branches for upper (blue) and lower (red) edges are 
indicated in different colors.   
  \begin{figure}[h]
   \psfrag{A}{\scriptsize(a)}
   \psfrag{BB}{\scriptsize(b)}
  \psfrag{K2}{ $K^\prime$}
   \psfrag{C}{ $C$}
 \psfrag{E1}{ \tiny$E_1$}
   \psfrag{E2}{\tiny $E_2$}
   \psfrag{E3}{\tiny $E_3$}
  \psfrag{E4}{\tiny $E_4$}
    \centering
  \includegraphics[width=250pt]{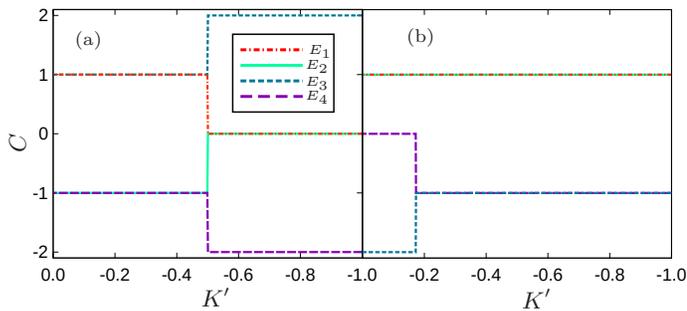}
  \caption{Variation of Chern numbers with $K^\prime$ for (a) 
$K=-0.5$, $J^\prime=-1$, and (b) $K=1$, $J^\prime=-0.5$. 
For each case $J=-1$, $D_{\rm m}=1$ and $h=1$.}
   \label{chern_no}
  \end{figure}  
  \begin{figure}[h]
  \psfrag{b}{ $\kappa_{xy} \hbar/k_{B}$}
 \psfrag{K2}{ $K^\prime$}
 \psfrag{A}{\tiny (a)}
   \psfrag{BB}{\tiny (b)}
   \psfrag{1}{\tiny $J=-1$}
   \psfrag{2}{\tiny $K=-0.5$}
   \psfrag{3}{\tiny $J^\prime=-1$}
   \psfrag{4}{\tiny $D_{\rm m}=1$}
   \psfrag{5}{\tiny $h=1$}
   \psfrag{6}{\tiny $k_{B}T=20$}
    \psfrag{7}{\tiny $K=-1$}
   \psfrag{8}{\tiny $J^\prime=-0.5$}
    \centering
  \includegraphics[width=250pt]{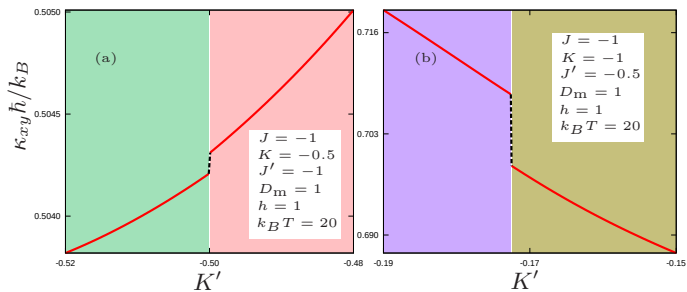}
  \caption{Variation of $\kappa_{xy} \hbar/k_{B}$ in the parameter space when $T$ is
fixed. Different regions are identified with distinct colors.}
   \label{Hall1}
  \end{figure}  
  \begin{figure}[h]
  \psfrag{b}{ $\kappa_{xy} \hbar/k_{B}$}  
   \psfrag{a}{ $k_B T$}
 \psfrag{A}{\tiny (a)}
   \psfrag{B}{\tiny (b)}
   \psfrag{C}{\tiny (c)}
   \psfrag{D}{\tiny (d)}
   \psfrag{E}{\tiny (e)}
    \psfrag{F}{\tiny (f)}
    \centering
  \includegraphics[width=200pt]{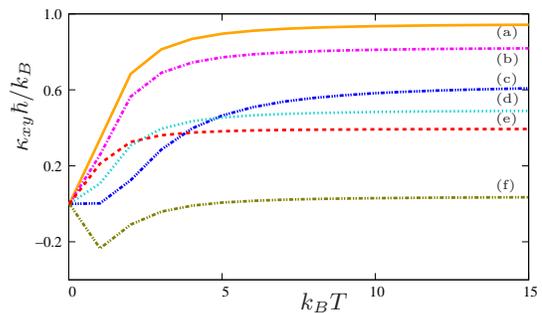}
  \caption{Variation of $\kappa_{xy}(T)$ with $T$ for 
a) $J=-1$, $D_{\rm m}=0.5$ for ${\mathcal C}=(11\bar 1\bar 1)$, 
(b) $J=-1$, $K'=-0.5$, $D_{\rm m}=0.5$ for ${\mathcal C}=(11\bar 20)$, 
(c) $J=-1$, $K=0.5$, $D_{\rm m}=0.5$ for ${\mathcal C}=(1\bar 11\bar 1)$, 
(d) $J=-1$, $J^\prime=-0.5$, $K'=-1$, $D_{\rm m}=1$ for ${\mathcal C}=(002\bar 2)$, 
(e)  $J^\prime=-0.5$, $K'=-0.5$, $D_{\rm m}=1$ for ${\mathcal C}=(\bar 222\bar 2)$, and 
(f) $J=-1$, $K=0.5$, $K'=-1$, $D_{\rm m}=1$ for ${\mathcal C}=(02\bar 20)$, 
with $h=1$. No value is assigned to those parameters when they are zero.}
   \label{Hall2}
  \end{figure}      
      
TMI phases as well as band gaps appear as soon as 
NNN DMI is switched on. Let us now describe the TMI phases 
shown in Fig \ref{energy3d}. Two distinct topological phases 
appear when all other  
NNN bond strengths are zero. These are 
${\mathcal C}=(11\bar 1\bar 1)$ and ${\mathcal C}=(1\bar 11\bar 1)$, 
as shown in Fig \ref{energy3d} (a) and (c),
respectively, when 
$J=-1$, $D_{\rm m}=0.5$ and  $J=-1$, $K=0.5$, $D_{\rm m}=0.5$, 
in two respective cases. 
${\mathcal C}$s are expressed following the 
ascending order of energy value in every case.  
The remaining four phases appear in the presence of 
other NNN bond strengths. Among them ${\mathcal C}=(\bar 222\bar 2)$ 
appears when all the NN interactions are absent. 
Therefore, TMIs with ${\mathcal C}=(11\bar 20)$, ${\mathcal C}=(002\bar 2)$ 
and ${\mathcal C}=(02\bar 20)$ emerges when 
both NN and NNN terms are present. However, 
appearance of those phases is by no means fixed for those 
particular values of the parameters. 
Those phases may appear for other combinations of parameter 
with different values also. 
But, no additional phase other that those six is by any means found to appear. 

Gapless edge states are shown in Fig \ref{edge}, where the number of 
edge states are found to satisfy the 
`bulk-edge correspondence' rule which states that 
  sum of the Chern number upto the $i$-th band, 
$\nu_i = \sum_{j\leqslant i}{\mathcal C}_j$, is equal to the 
number of pair of edge states in the gap \cite{Mook}. 
Which means that the values of the Chern numbers can be derived, 
otherwise, from the edge state pattern itself.

Topological phase transition (TPT) may be noted in the parameter space 
upon changing the values of the parameters. One such transition occurs 
when $K'$ becomes non-zero but $J=-1$ and $D_{\rm m}=0.5$. 
In this case, the system undergoes a transition from the 
state $(11\bar 1\bar 1)$ to another state $(11\bar 20)$. 
Distribution of ${\mathcal C}$s of those two phases 
around the transition point can be understood in the following way. 
Gap between the upper two bands vanishes at the transition 
point in the parameter space due to the presence of a Dirac 
cone at the band touching point. 
When the gap reopens ${\mathcal C}$s of the respective upper two bands 
change by $\pm 1$, resulting in the redistribution of them. 
Occurrence of other TPTs may be explained in similar fashion. 
For example,  transition from $(11\bar 1\bar 1)$ to $(1\bar 11\bar 1)$ 
takes place by switching on the $K$. 
In this case, two intermediate bands touch in such a way that a 
Dirac cone is formed at the band touching point. 

Another pair of TPT is shown in Fig \ref{chern_no}, 
where the appearance of topological phases is noted 
with the variation of $K'$. The energies, $E_1$, $E_2$, $E_3$ and $E_4$ 
are denoted according to the ascending order of their values.
 Fig \ref{chern_no} (a) shows that two nontrivial topological phases, 
$(1\bar 11\bar 1)$ and $(002\bar 2)$
appear around $K'=-0.5$. 
TMI with $(11\bar 1\bar 1)$ is found when $K'<-0.16$, as shown in Fig \ref{chern_no} (b).
Upon increase of $K'$, Chern numbers of the lower two bands remain unchanged while 
those of upper two bands are exchanged by $\pm1$ leading to a
new topological phase with $(11\bar 20)$. 
Transition between the same set of topological phases may take place 
in a variety of ways. 
Variation of $\kappa_{xy}$ with respect to $K'$ 
for four different TMI phases is shown in Fig \ref{Hall1} with different colors.  
Those are plotted when $k_{B}T=20$. Sudden jump in $\kappa_{xy}$ corresponds to 
the point where TPT occurs. 
Similarly, variation of $\kappa_{xy}$ with respect to $T$ 
is shown in Fig \ref{Hall2}.       
   \begin{figure}[t]
 \psfrag{E}{ $E$}
 \psfrag{kx}{\tiny $k_x$}
 \psfrag{ky}{\tiny $k_y$}
   \psfrag{AA}{(a)}
   \psfrag{c1}{\tiny ${\mathcal C}_1=0$}
\psfrag{c2}{\tiny ${\mathcal C}_2=1$}
\psfrag{c3}{\tiny ${\mathcal C}_3=\bar 1$}
\psfrag{c4}{\tiny ${\mathcal C}_4=0$}
\begin{minipage}{0.18\textwidth}
  \includegraphics[width=170pt]{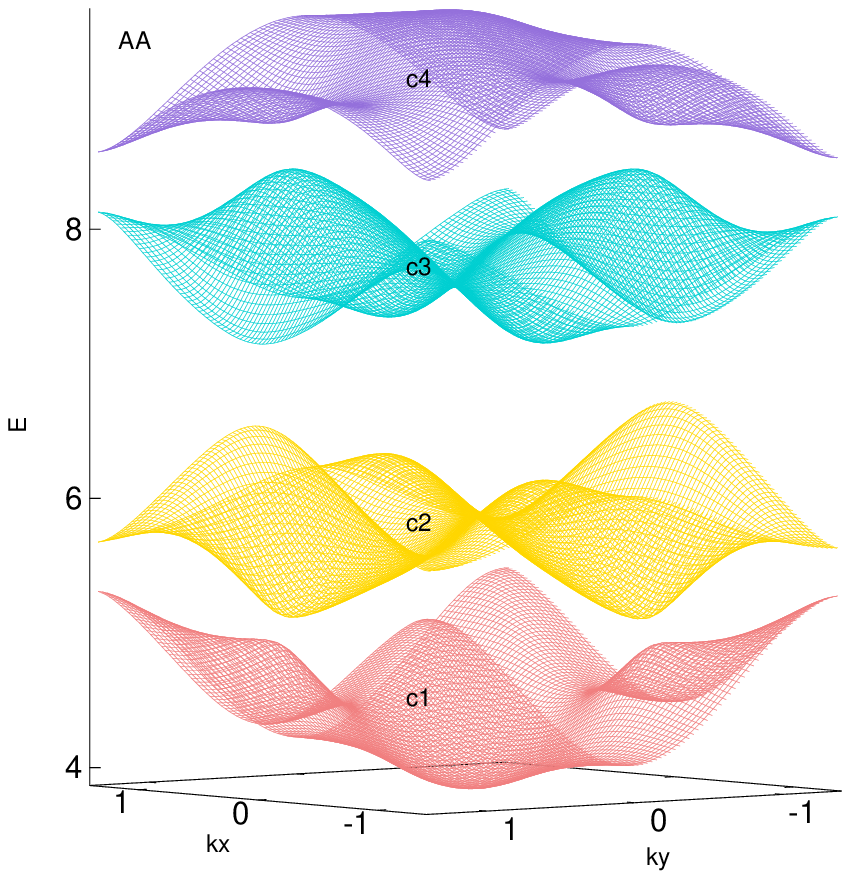}
  \end{minipage}\hfill   
  \begin{minipage}{0.28\textwidth}
  \psfrag{E}{}
  \psfrag{AA}{(b)}
    \psfrag{c1}{\tiny ${\mathcal C}_1=1$}
\psfrag{c2}{\tiny ${\mathcal C}_2=0$}
\psfrag{c3}{\tiny ${\mathcal C}_3=0$}
\psfrag{c4}{\tiny ${\mathcal C}_4=\bar 1$}
    \centering
  \includegraphics[width=170pt]{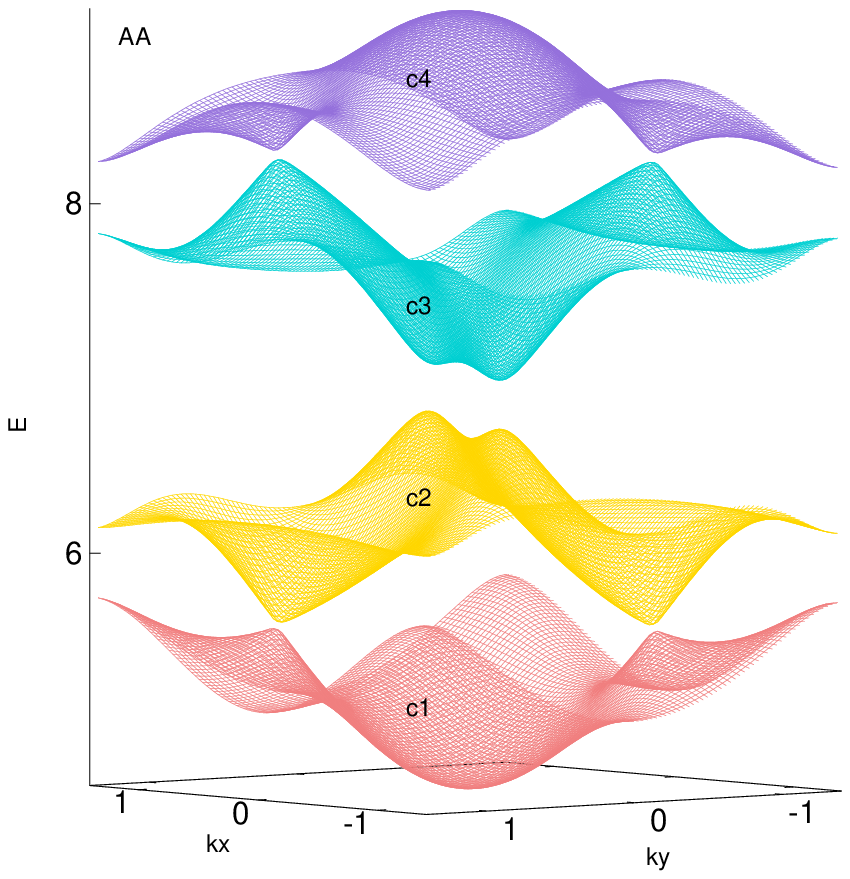}
  \end{minipage}\hfill 
  
  \begin{minipage}{0.18\textwidth}
  \psfrag{E}{ $E$}
\psfrag{AA}{(c)}
   \psfrag{c1}{\tiny ${\mathcal C}_1=2$}
\psfrag{c2}{\tiny ${\mathcal C}_2=\bar 3$}
\psfrag{c3}{\tiny ${\mathcal C}_3=3$}
\psfrag{c4}{\tiny ${\mathcal C}_4=\bar 2$}
    \centering
  \includegraphics[width=170pt]{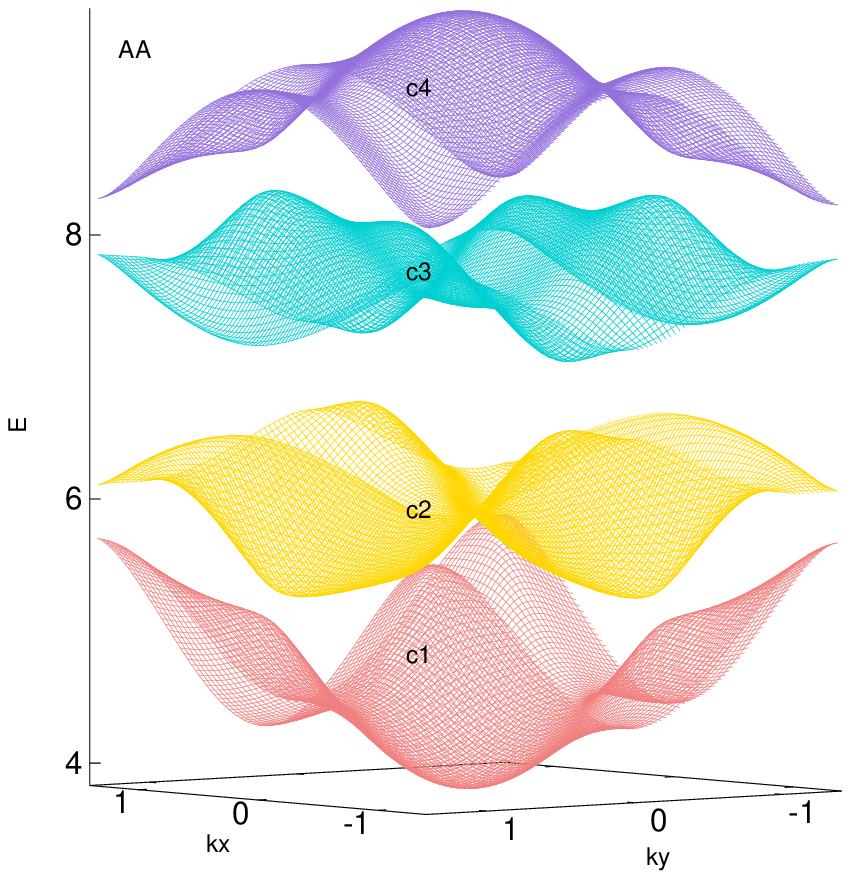}
   \end{minipage}\hfill   
  \begin{minipage}{0.28\textwidth}
\psfrag{E}{}
\psfrag{AA}{(d)}
    \psfrag{c1}{\tiny ${\mathcal C}_1=1$}
\psfrag{c2}{\tiny ${\mathcal C}_2=\bar 2$}
\psfrag{c3}{\tiny ${\mathcal C}_3=2$}
\psfrag{c4}{\tiny ${\mathcal C}_4=\bar 1$}
    \centering
  \includegraphics[width=170pt]{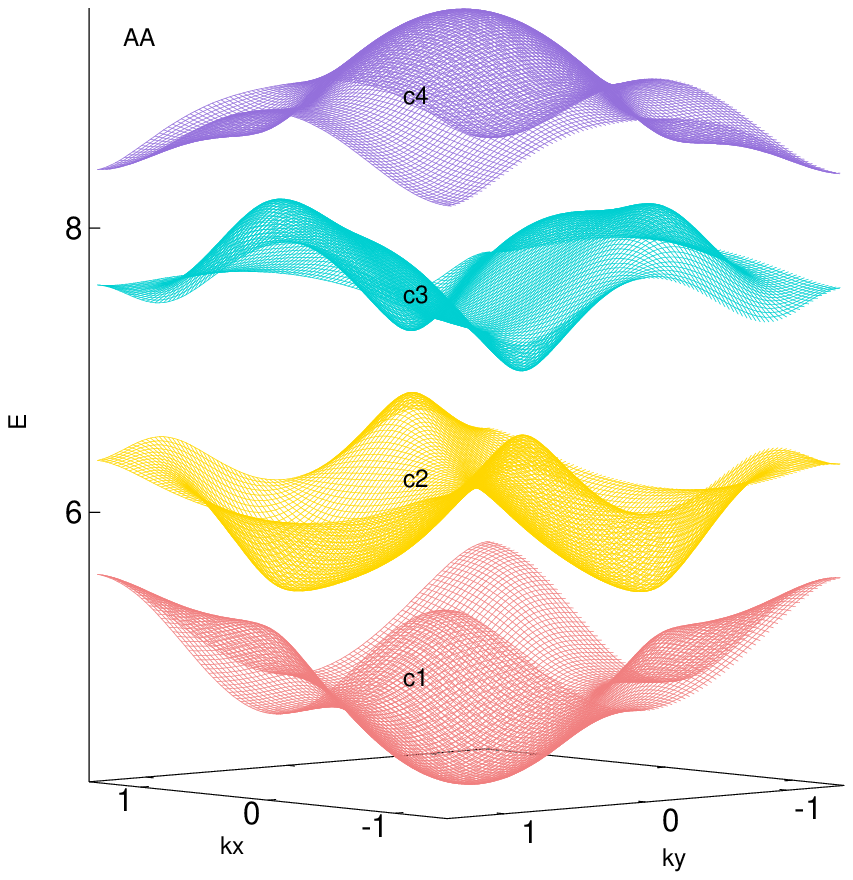}
  \end{minipage}\hfill 

 \begin{minipage}{0.18\textwidth}
\psfrag{E}{ $E$}
 \psfrag{kx}{\tiny $k_x$}
 \psfrag{ky}{\tiny $k_y$}
   \psfrag{AA}{(e)}
   \psfrag{c1}{\tiny ${\mathcal C}_1=1$}
\psfrag{c2}{\tiny ${\mathcal C}_2=\bar 1$}
\psfrag{c3}{\tiny ${\mathcal C}_3=1$}
\psfrag{c4}{\tiny ${\mathcal C}_4=\bar 1$}
  \includegraphics[width=170pt]{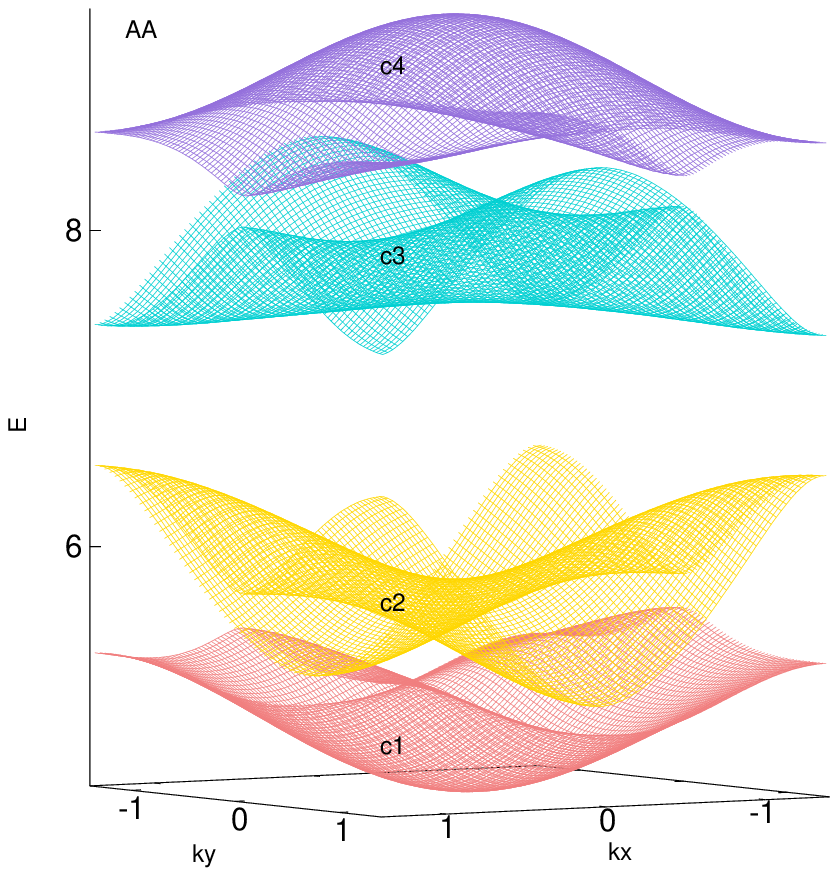}
  \end{minipage}\hfill   
  \begin{minipage}{0.28\textwidth}
  \psfrag{E}{}
\psfrag{AA}{(f)}
    \psfrag{c1}{\tiny ${\mathcal C}_1=2$}
\psfrag{c2}{\tiny ${\mathcal C}_2=\bar 2$}
\psfrag{c3}{\tiny ${\mathcal C}_3=2$}
\psfrag{c4}{\tiny ${\mathcal C}_4=\bar 2$}
    \centering
  \includegraphics[width=170pt]{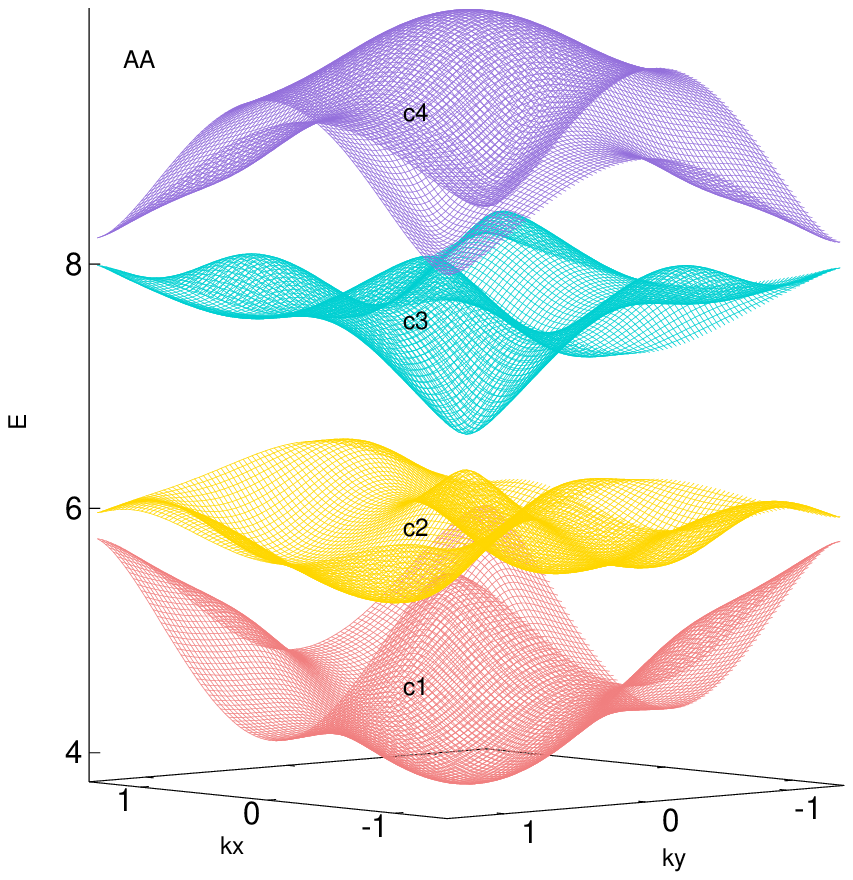}
  \end{minipage}\hfill   
  \caption{The magnon bands.
 Values of the parameters are:  
(a) $K_x=-0.8$, $K_y=-0.6$, $K^\prime_x=-1$, $K^\prime_y=0.4$ for ${\mathcal C}=(01\bar 10)$, 
(b) $K_x=-0.7$, $K_y=-0.2$, $K^\prime_x=-0.5$, $K^\prime_y=0.1$ for ${\mathcal C}=(100\bar 1)$, 
(c) $K_x=-0.8$, $K^\prime_x=-1$, $K^\prime_y=-0.4$ for ${\mathcal C}=(2\bar 33\bar 2)$,
(d) $K_x=-1$, $K_y=-0.1$, $K^\prime_x=-0.6$, $K^\prime_y=-0.3$ for ${\mathcal C}=(1\bar 22\bar 1)$, 
(e) $K_x=-0.7$, $K_y=-0.8$, $K^\prime_x=-0.1$, $K^\prime_y=-0.5$ for 
${\mathcal C}=(1\bar 11\bar 1)$, and
(f) $K_x=-0.7$, $K_y=-0.4$, $K^\prime_x=-0.8$, $K^\prime_y=-0.9$ for 
${\mathcal C}=(2\bar 22\bar 2)$, with $K_z=-2$, $K^\prime_z=-1$, $D_{\rm m}=1$, $h=1$.
No value is assigned to those parameters when they are zero.}
   \label{energy3dnew}
  \end{figure}      
  
   \begin{figure}[h]
  \psfrag{E}{ $E$}
\psfrag{X}{X}
\psfrag{G}{  $\Gamma$}
\psfrag{K}{ $\bold{k}$}
\psfrag{M}{ M}   
   \psfrag{AA}{\scriptsize(a)}
   \begin{minipage}{0.21\textwidth}
   \psfrag{c1}{\tiny ${\mathcal C}_1=0$}
\psfrag{c2}{\tiny ${\mathcal C}_2=1$}
\psfrag{c3}{\tiny ${\mathcal C}_3=\bar 1$}
\psfrag{c4}{\tiny ${\mathcal C}_4=0$}
    \centering
  \includegraphics[width=125pt]{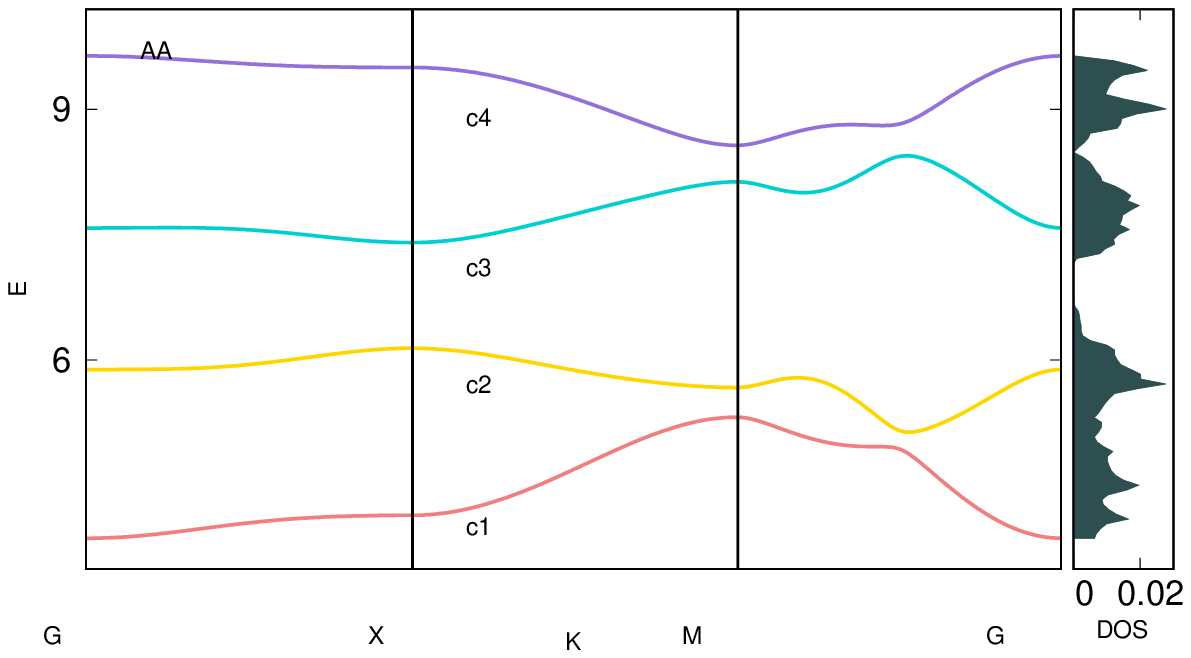}
  \end{minipage}\hfill   
  \begin{minipage}{0.25\textwidth}
  \psfrag{E}{ }
   \psfrag{AA}{\scriptsize(b)}
    \psfrag{c1}{\tiny ${\mathcal C}_1=1$}
\psfrag{c2}{\tiny ${\mathcal C}_2=0$}
\psfrag{c3}{\tiny ${\mathcal C}_3=0$}
\psfrag{c4}{\tiny ${\mathcal C}_4=\bar 1$}
     \centering
  \includegraphics[width=125pt]{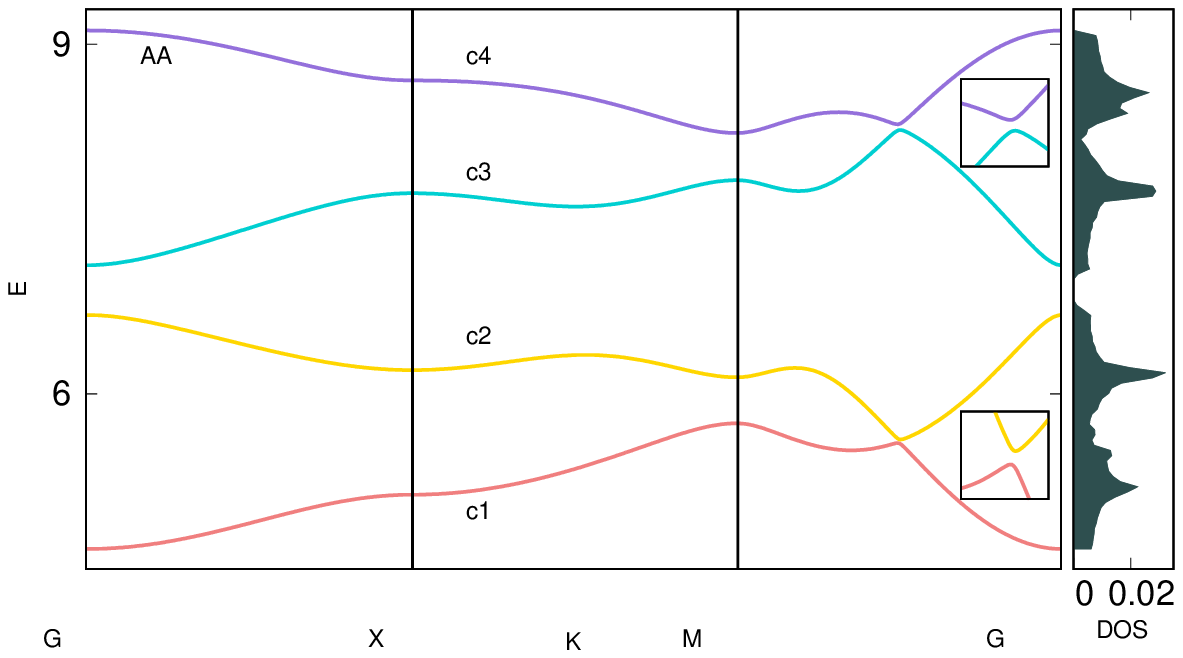}
   \end{minipage}\hfill   
    \begin{minipage}{0.21\textwidth}
    \psfrag{E}{ $E$}
   \psfrag{AA}{\scriptsize(c)}
  \psfrag{c1}{\tiny ${\mathcal C}_1=2$}
\psfrag{c2}{\tiny ${\mathcal C}_2=\bar 3$}
\psfrag{c3}{\tiny ${\mathcal C}_3=3$}
\psfrag{c4}{\tiny ${\mathcal C}_4=\bar 2$}
    \centering
  \includegraphics[width=125pt]{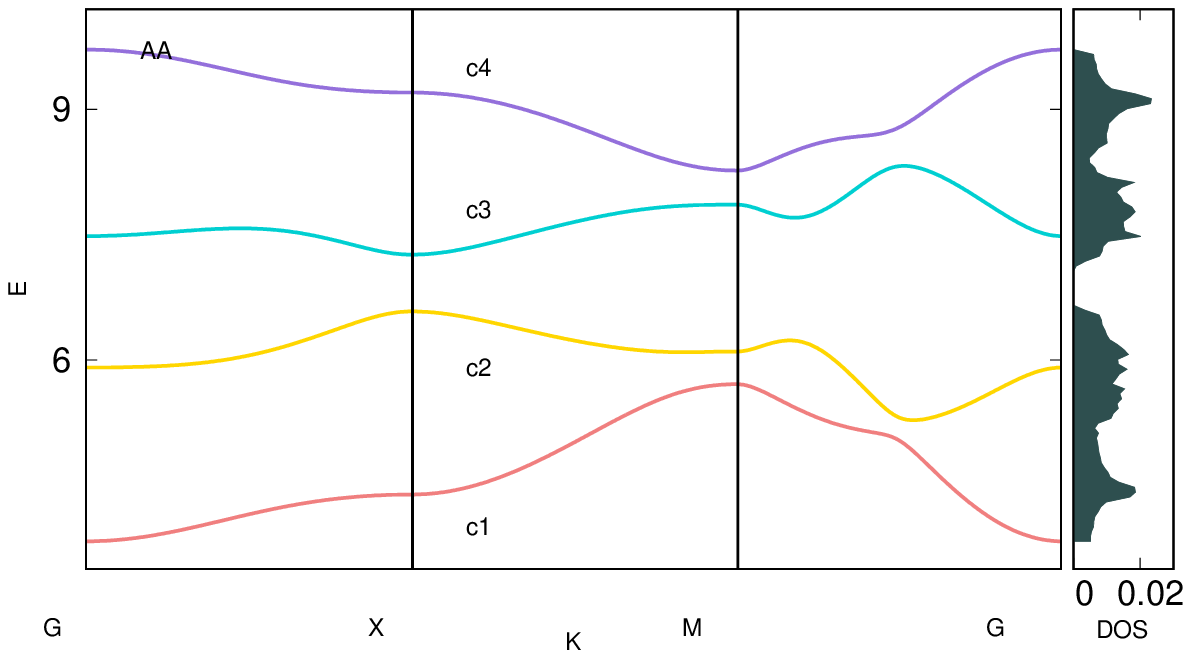}
   \end{minipage}\hfill   
  \begin{minipage}{0.25\textwidth}
  \psfrag{E}{ }
 \psfrag{AA}{\scriptsize(d)}
   \psfrag{c1}{\tiny ${\mathcal C}_1=1$}
\psfrag{c2}{\tiny ${\mathcal C}_2=\bar 2$}
\psfrag{c3}{\tiny ${\mathcal C}_3=2$}
\psfrag{c4}{\tiny ${\mathcal C}_4=\bar 1$}
     \centering
  \includegraphics[width=125pt]{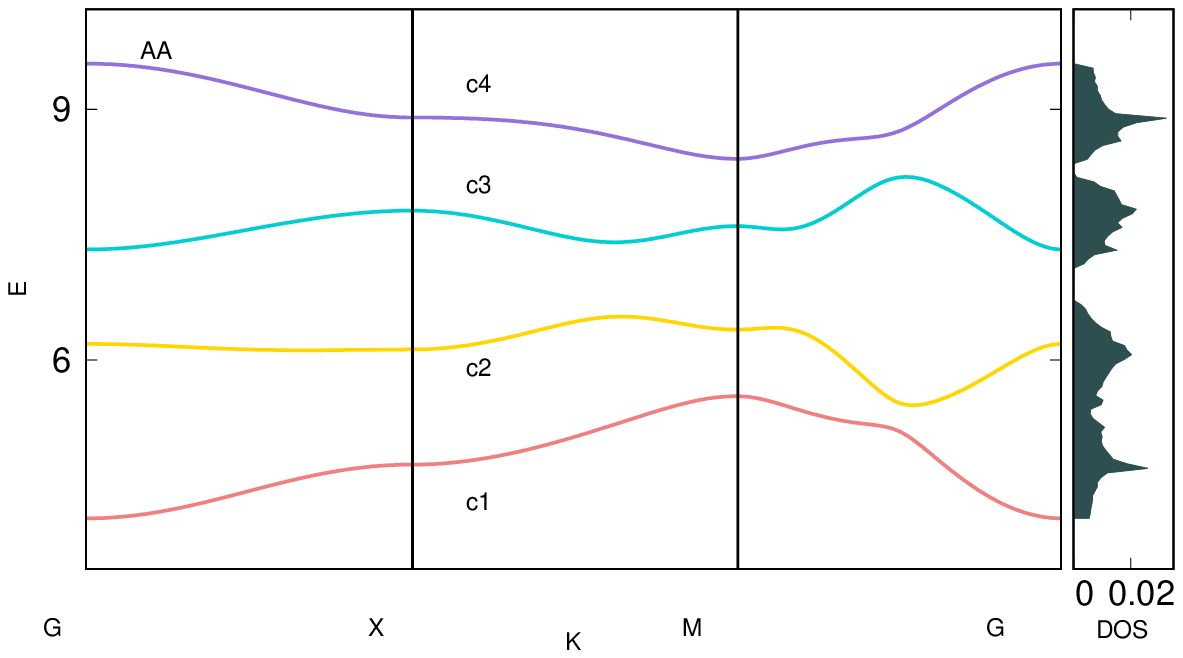}
  \end{minipage}\hfill   
    \begin{minipage}{0.21\textwidth}
    \psfrag{E}{ $E$}
    \psfrag{AA}{(e)}
   \psfrag{c1}{\tiny ${\mathcal C}_1=1$}
\psfrag{c2}{\tiny ${\mathcal C}_2=\bar 1$}
\psfrag{c3}{\tiny ${\mathcal C}_3=1$}
\psfrag{c4}{\tiny ${\mathcal C}_4=\bar 1$}
 \includegraphics[width=125pt]{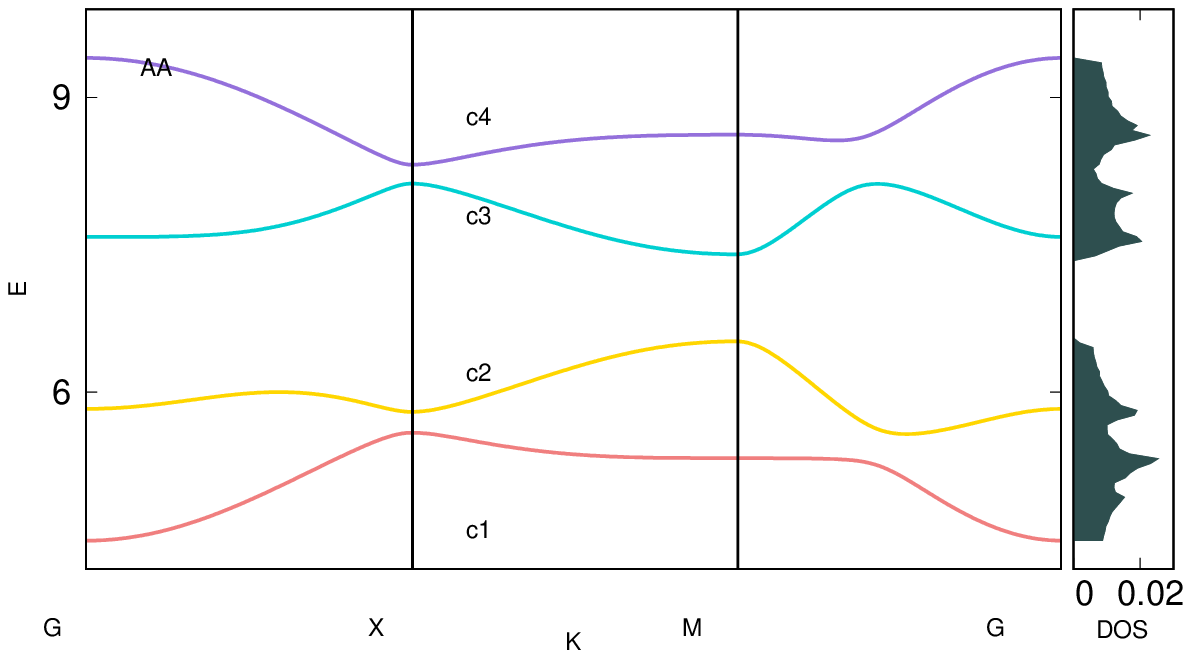}
 \end{minipage}\hfill   
  \begin{minipage}{0.25\textwidth}
  \psfrag{E}{ }
 \psfrag{AA}{(f)}
   \psfrag{c1}{\tiny ${\mathcal C}_1=2$}
\psfrag{c2}{\tiny ${\mathcal C}_2=\bar 2$}
\psfrag{c3}{\tiny ${\mathcal C}_3=2$}
\psfrag{c4}{\tiny ${\mathcal C}_4=\bar 2$}
 \includegraphics[width=125pt]{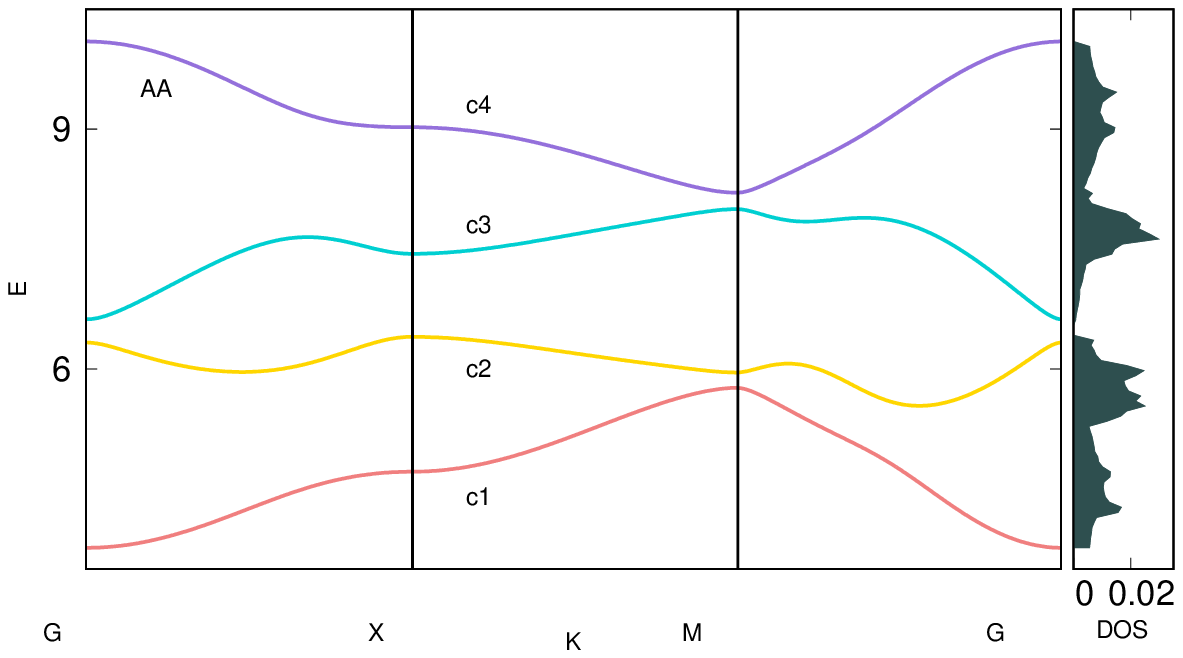}
 \end{minipage}\hfill   
  \caption{Dispersion relation along the high-symmetry points of Brillouin zone. 
The side panel shows the DOS. Values of the parameters are the 
same as Fig \ref{energy3dnew} for the respective plots.}
     \label{energybznew}
  \end{figure}     
            
\subsection{Anisotropic Kitaev coupling}    
Anisotropic Kitaev coupling corresponds to $K_x\neq K_y \neq K_z$ 
as well as $K'_x\neq K'_y \neq K'_z$. 
Additional conditions, $J=0$ and $J'=0$, are assumed to make the
system closer to the Kitaev model. In this limit, system
only holds the Kitaev terms apart from the DMI.
Kitaev model with anisotropic NN coupling, ($K_x\neq K_y \neq K_z$), 
on both the honeycomb and CaVO lattices 
have been solved exactly \cite{Kitaev,Sun}. 
Both the systems host gapless and gapped phases in the ground state 
phase diagram. Identically, both the systems exhibit a unique 
topological phase, ${\mathcal C}$=($1\bar 1$), in the presence of 
magnetic field, when they are studied in terms of Majorana fermion.  
     
\begin{figure}[t]
 \psfrag{E}{$E$}
 \psfrag{qx}{\scriptsize $k_x$}
 \psfrag{w}{ \scriptsize$|\psi|^2$}
 \psfrag{AA}{\scriptsize(a)}
 \psfrag{site}{\scriptsize site}
 \psfrag{ub}{\tiny upper edge}
 \psfrag{lb}{\tiny lower edge}
   \begin{minipage}{0.21\textwidth}
   \includegraphics[width=130pt]{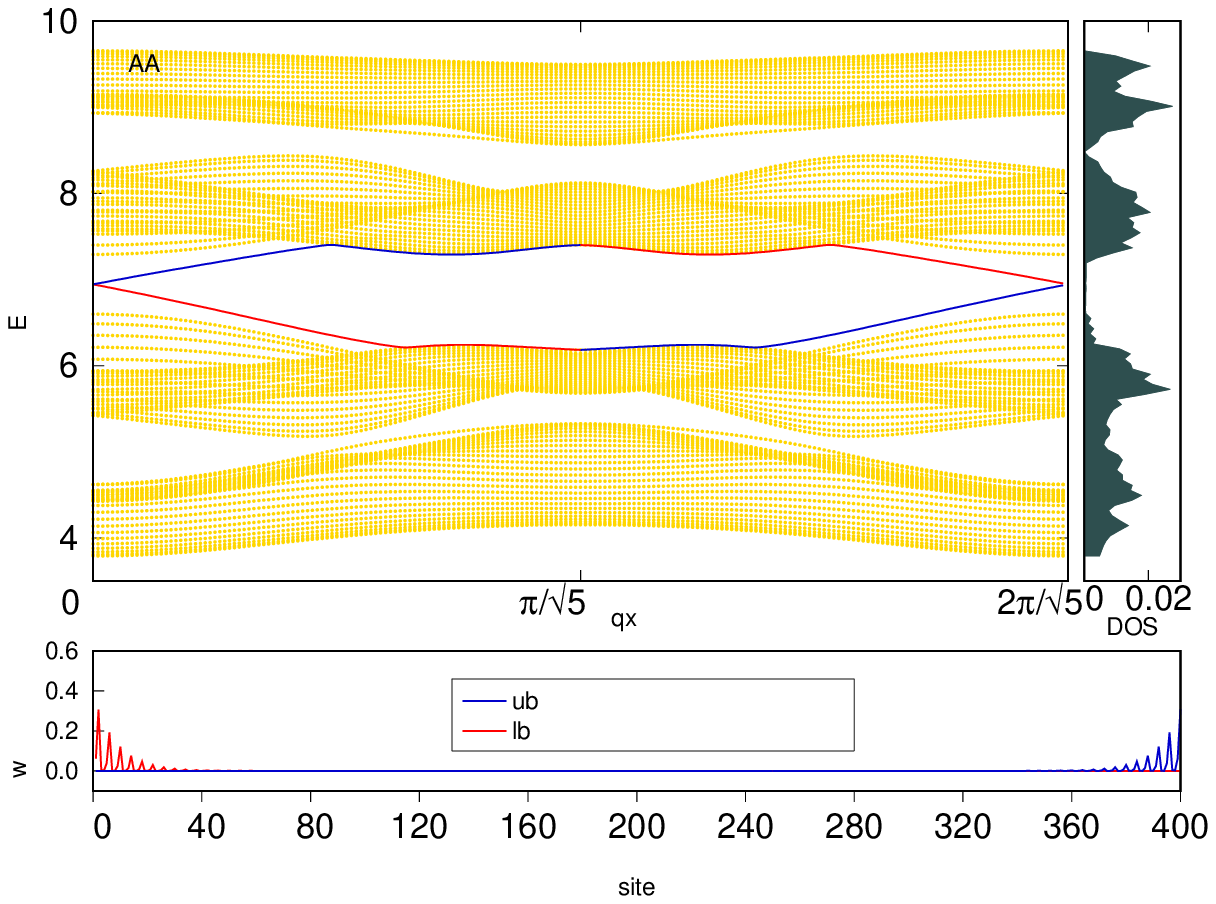}
    \end{minipage}\hfill
     \psfrag{E}{}
   \psfrag{w}{}
    \psfrag{AA}{\scriptsize(b)}
   \begin{minipage}{0.24\textwidth}
   \includegraphics[width=130pt]{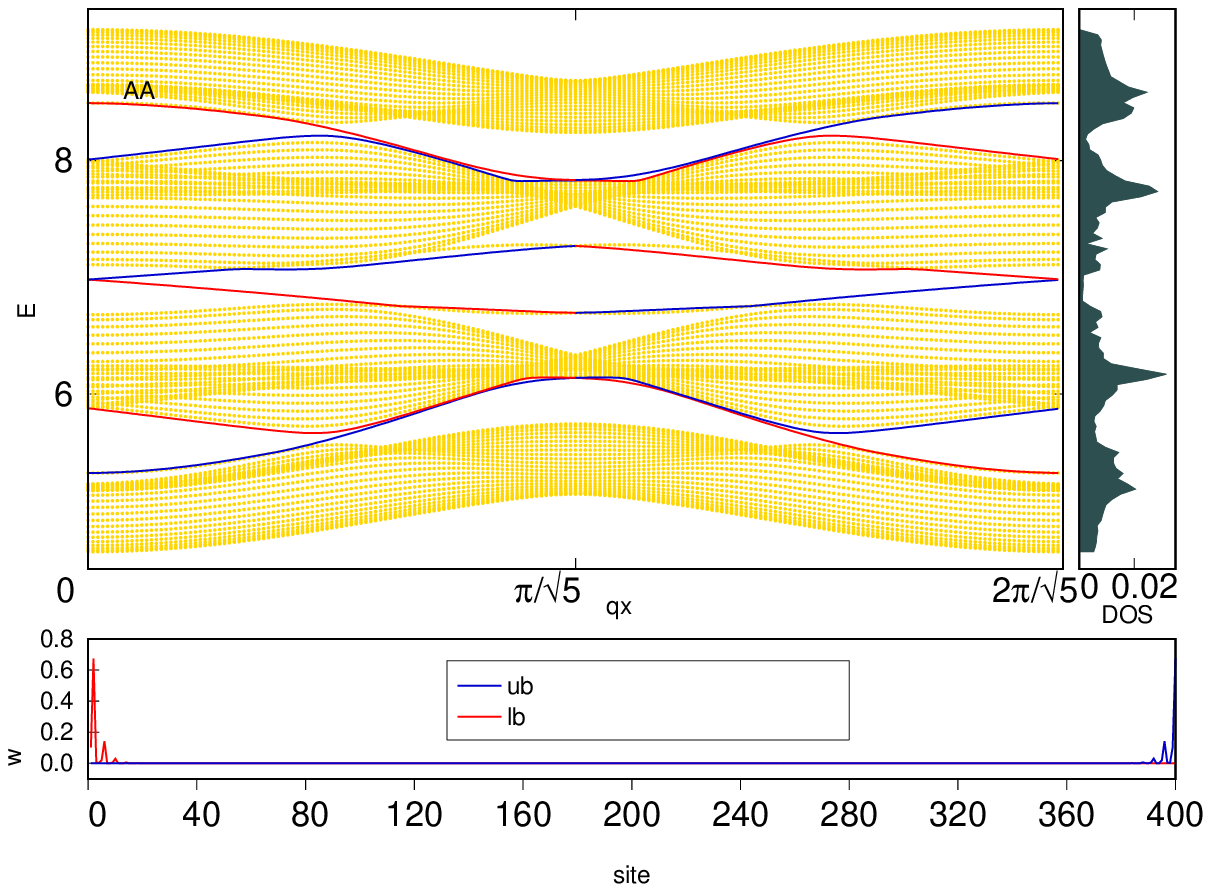}
   \end{minipage}\hfill
     \vskip 0.07cm
   \begin{minipage}{0.21\textwidth}
     \psfrag{E}{ $E$}
   \psfrag{w}{\scriptsize $|\psi|^2$}   
\psfrag{AA}{\scriptsize(c)}
   \includegraphics[width=130pt]{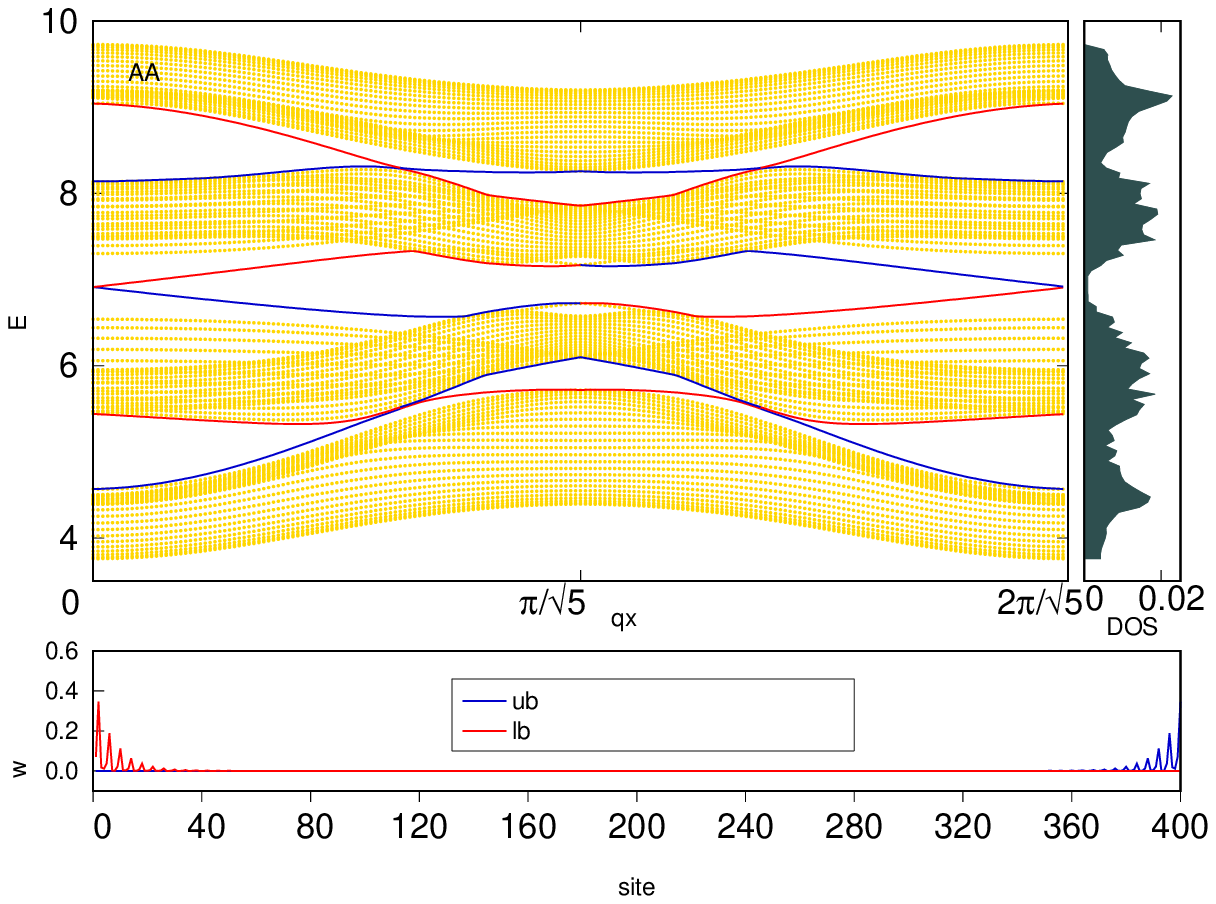}
 \end{minipage}\hfill
   \begin{minipage}{0.24\textwidth}
    \psfrag{E}{}
   \psfrag{w}{}   
 \psfrag{AA}{\scriptsize(d)}
   \includegraphics[width=130pt]{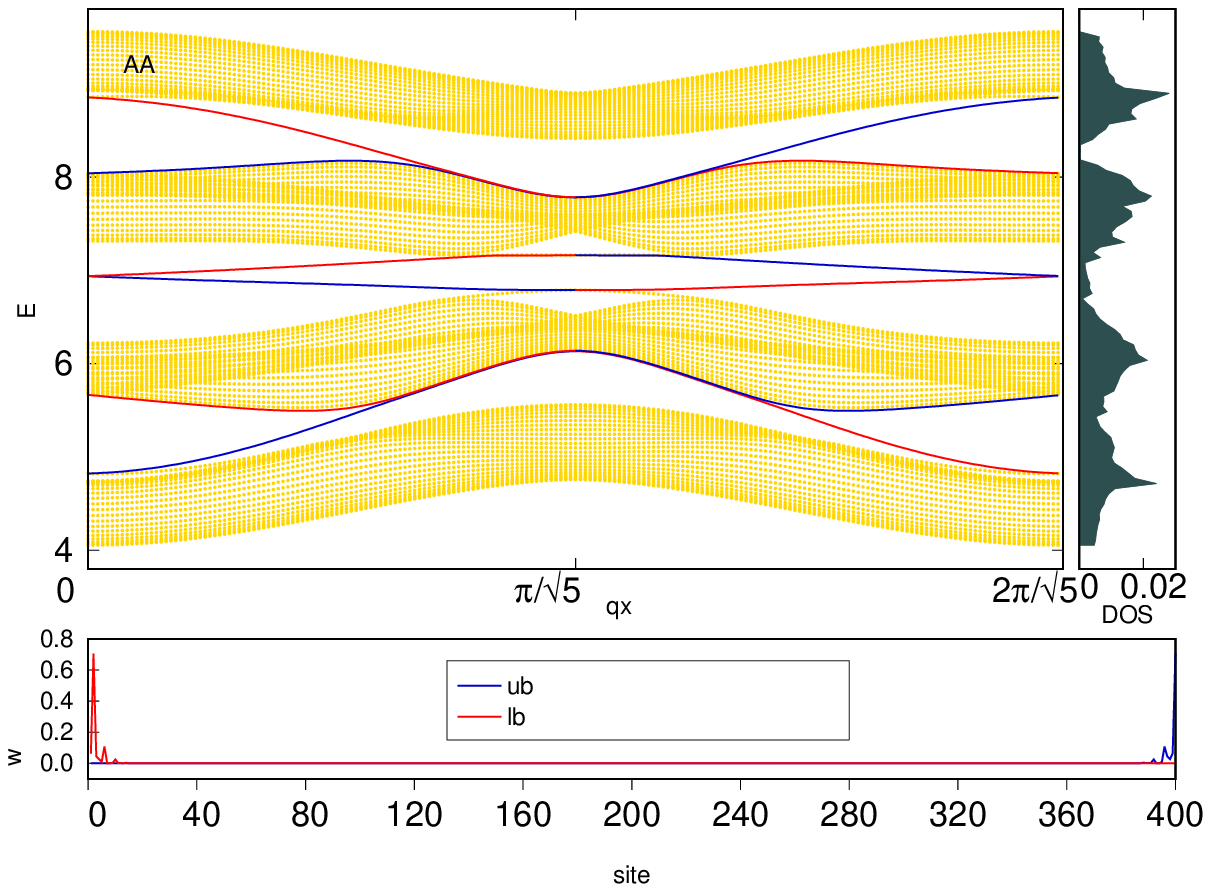}   
 \end{minipage}\hfill
   \vskip 0.07cm
   \begin{minipage}{0.21\textwidth}
     \psfrag{E}{ $E$}
   \psfrag{w}{\scriptsize $|\psi|^2$}
 \psfrag{AA}{\scriptsize(e)}
   \includegraphics[width=130pt]{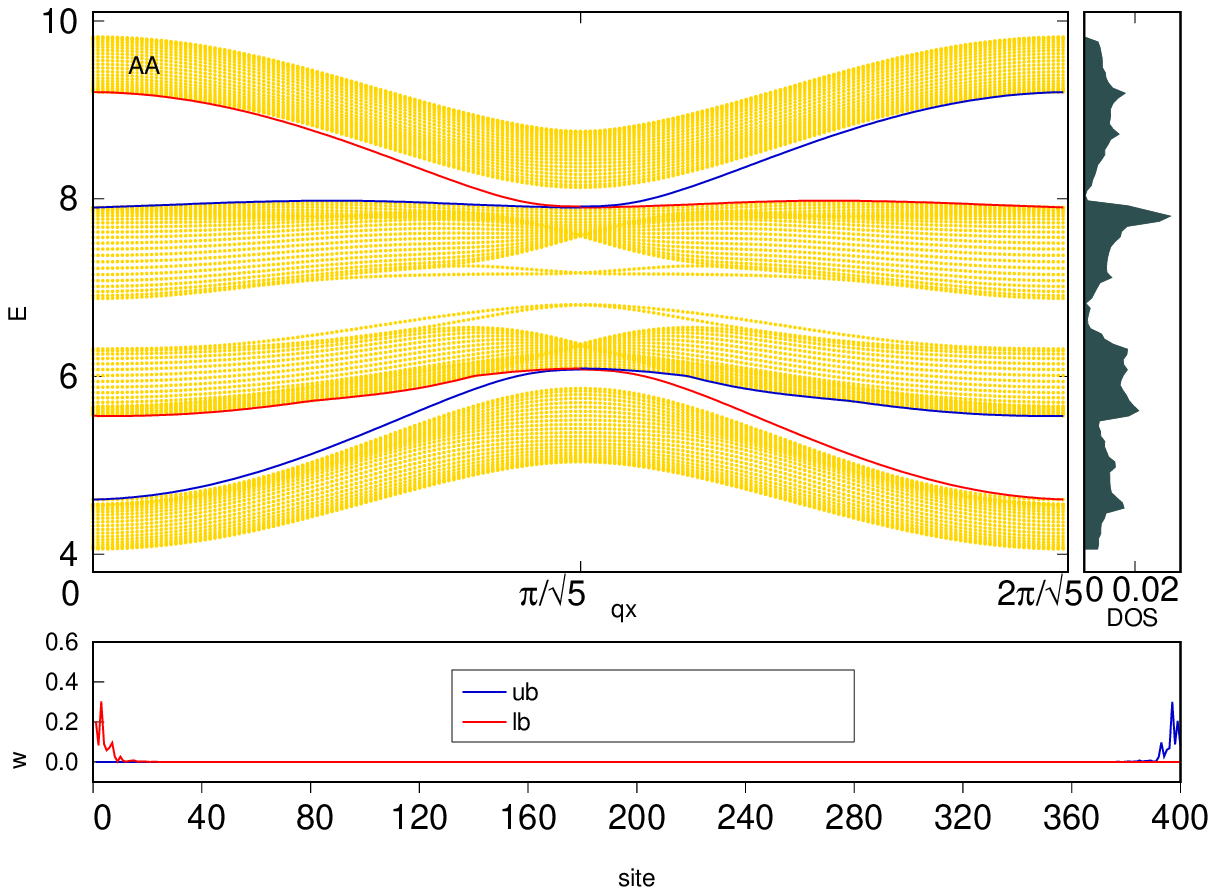}
 \end{minipage}\hfill
   \begin{minipage}{0.24\textwidth}
    \psfrag{E}{}
   \psfrag{w}{}
 \psfrag{AA}{\scriptsize(f)}
   \includegraphics[width=130pt]{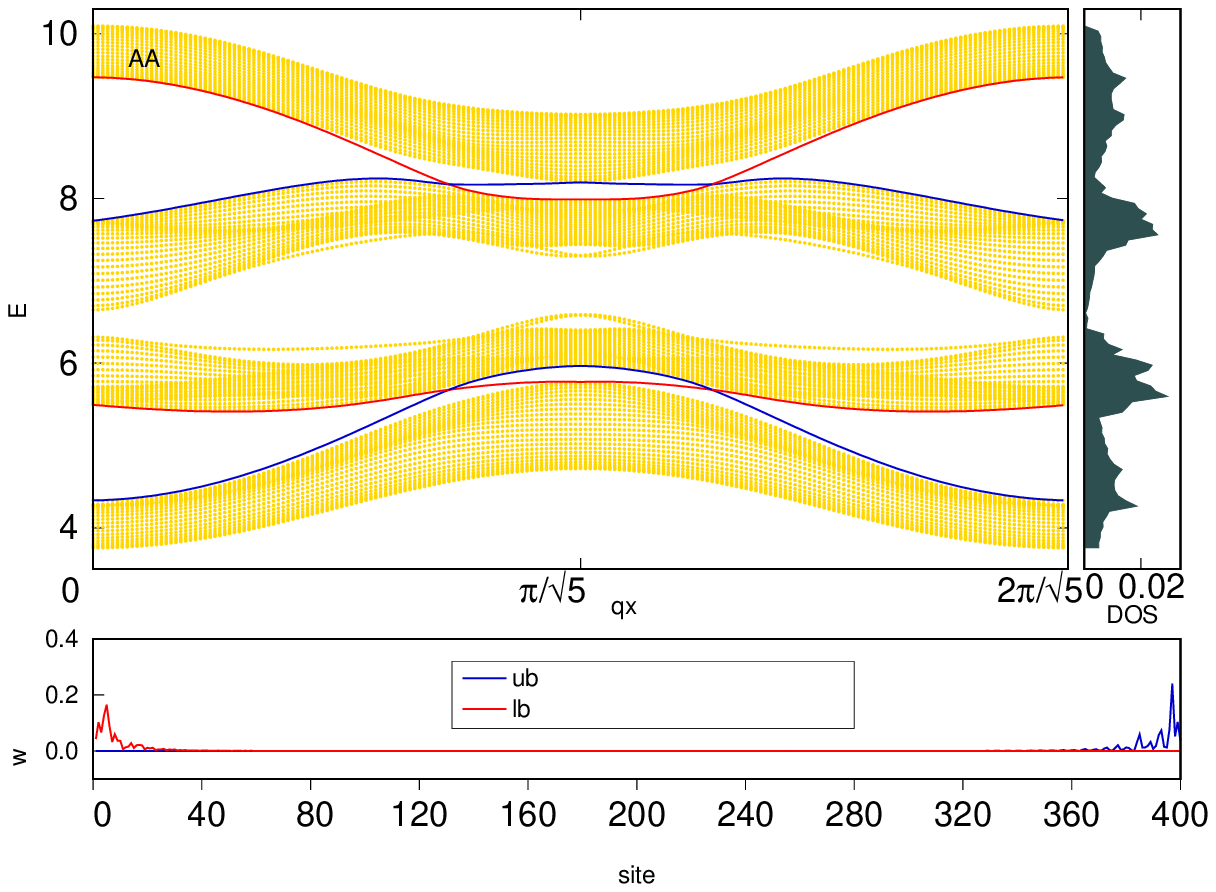}   
 \end{minipage}\hfill
  \caption{Magnon dispersions of bulk-edge states in the one-dimensional BZ. 
Upper and lower edge modes are drawn in blue and red lines, respectively, while 
bulk modes are in golden points.
The side panel shows the density of edge states.
 The lower panel indicates variation of probability density of both edge
modes with respect to site number for a fixed $k_x$. Values of the parameters are:  
(a) $K_x=-0.8$, $K_y=-0.6$, $K^\prime_x=-1$, $K^\prime_y=0.4$ for ${\mathcal C}=(01\bar 10)$, 
(b) $K_x=-0.7$, $K_y=-0.2$, $K^\prime_x=-0.5$, $K^\prime_y=0.1$ for ${\mathcal C}=(100\bar 1)$, 
(c) $K_x=-0.8$, $K^\prime_x=-1$, $K^\prime_y=-0.4$ for ${\mathcal C}=(2\bar 33\bar 2)$,
(d) $K_x=-1$, $K_y=-0.1$, $K^\prime_x=-0.6$, $K^\prime_y=-0.3$ for ${\mathcal C}=(1\bar 22\bar 1)$, 
(e) $K_x=-0.7$, $K_y=-0.8$, $K^\prime_x=-0.1$, $K^\prime_y=-0.5$ for 
${\mathcal C}=(1\bar 11\bar 1)$, and
(f) $K_x=-0.7$, $K_y=-0.4$, $K^\prime_x=-0.8$, $K^\prime_y=-0.9$ 
for ${\mathcal C}=(2\bar 22\bar 2)$, with $K_z=-2$, $K^\prime_z=-1$, $D_{\rm m}=1$, $h=1$.
No value is assigned to those parameters when they are zero.}
 \label{edgenew}
   \end{figure}

\begin{figure}[h]
   \psfrag{AA}{\scriptsize (a)}
   \psfrag{k1}{ $K_y$}
  \psfrag{k2}{ $K_y^\prime$}
  \psfrag{c1}{\scriptsize $(01\bar 10)$}
    \psfrag{c2}{ \scriptsize $(0\bar 110)$}
     \psfrag{c5}{\scriptsize $(2\bar 33\bar 2)$}
      \psfrag{c6}{\scriptsize  $(\bar 23\bar 32)$}
       \psfrag{c11}{\scriptsize $(2\bar 22\bar 2)$}
        \psfrag{c12}{\scriptsize $(\bar 22\bar 22)$}
   \psfrag{BB}{\scriptsize (b)}
  \psfrag{c3}{\scriptsize $(100\bar 1)$}
    \psfrag{c4}{ \scriptsize $(\bar 1001)$}
     \psfrag{c7}{\scriptsize $(1\bar 22\bar 1)$}
      \psfrag{c8}{\scriptsize  $(\bar 12\bar 21)$}
       \psfrag{c9}{\scriptsize $(1\bar 11\bar 1)$}
        \psfrag{c10}{\scriptsize  $(\bar 11\bar 11)$}
         \psfrag{C13}{\scriptsize \color{white} $(0000)$}
  \includegraphics[width=250pt]{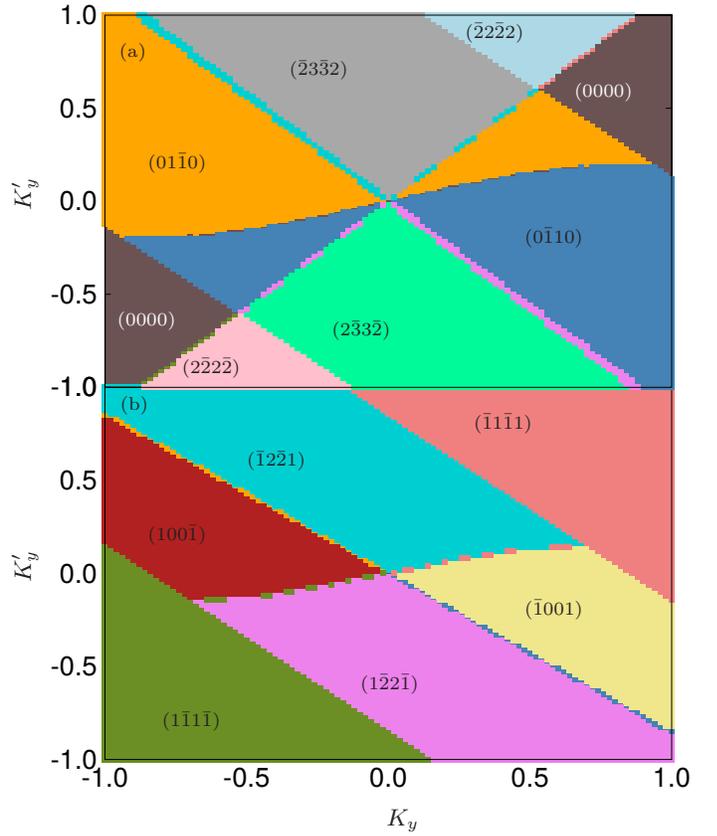}
  \caption{Regions of TMI phases of the system in $K_y$-$K_y^\prime$ 
parameter space for (a) $K_x^\prime=-0.8$, and (b) $K_x^\prime=-0.6$, 
with $K_x=-0.7$, $K_z=-2$, $K^\prime_z=-1$, $D_{\rm m}=1$, $h=1$. Trivial region is 
indicated by $(0000)$.}
   \label{chern_no_new}
  \end{figure}

\begin{figure}[h]
  \psfrag{b}{\scriptsize $\kappa_{xy} \hbar/k_{B}$}
 \psfrag{K1}{\scriptsize {$K_y$}}
 \psfrag{K2}{\scriptsize $K'_y$}
 \psfrag{A}{\scriptsize (a)}
   \psfrag{B}{\scriptsize (b)}
   \psfrag{C}{\scriptsize (c)}
   \psfrag{D}{\scriptsize (d)}
   \psfrag{E}{\scriptsize (e)}
   \psfrag{F}{\scriptsize (f)}
    \psfrag{1}{\tiny {$K_x^\prime=-0.8$}}
     \psfrag{2}{\tiny $K_y^\prime=0.85$}
     \psfrag{3}{\tiny $K_y=-0.5$}
     \psfrag{4}{\tiny $K_y^\prime=-0.85$}
    \psfrag{5}{\tiny {$K_x^\prime=-0.6$}}
     \psfrag{6}{\tiny $K_y=-0.8$}
       \psfrag{7}{\tiny $K_y=0$}
      \psfrag{8}{\tiny $K_y=0.8$}
    \centering
  \includegraphics[width=250pt]{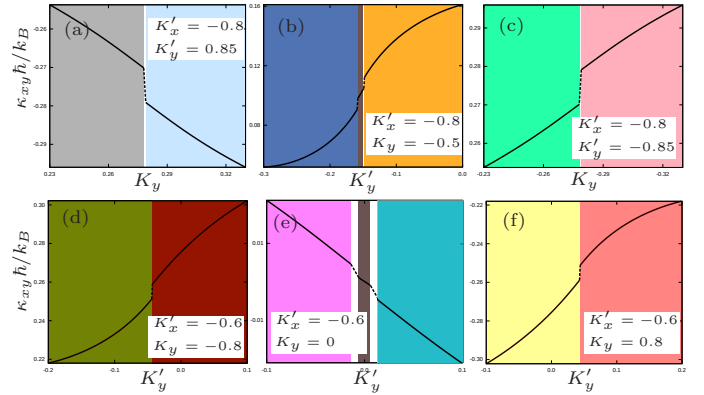}
  \caption{Variation of $\kappa_{xy} \hbar/k_{B}$ in the parameter space when 
$K_x=-0.7$, $K_z=-2$, $K'_z=-1$, $D_{\rm m}=1$, $h=1$, $k_B T=20$. Different 
regions are identified with distinct colors,
those are used before in Fig \ref{chern_no_new}.}
   \label{Hallnew1}
  \end{figure}

\begin{figure}[h]
  \psfrag{AA}{ (I)}
  \psfrag{BB}{ (II)}
  \psfrag{b}{ $\kappa_{xy} \hbar/k_{B}$}
   \psfrag{a}{ $k_B T$}
 \psfrag{A}{\tiny (a)}
   \psfrag{B}{\tiny (b)}
   \psfrag{C}{\tiny (c)}
   \psfrag{D}{\tiny (d)}
   \psfrag{E}{\tiny (e)}
   \psfrag{F}{\tiny (f)}
   \psfrag{1}{\tiny (g)}
   \psfrag{2}{\tiny (h)}
   \psfrag{3}{\tiny (i)}
   \psfrag{4}{\tiny (j)}
   \psfrag{5}{\tiny (k)}
   \psfrag{6}{\tiny (l)}
    \centering
  \includegraphics[width=200pt]{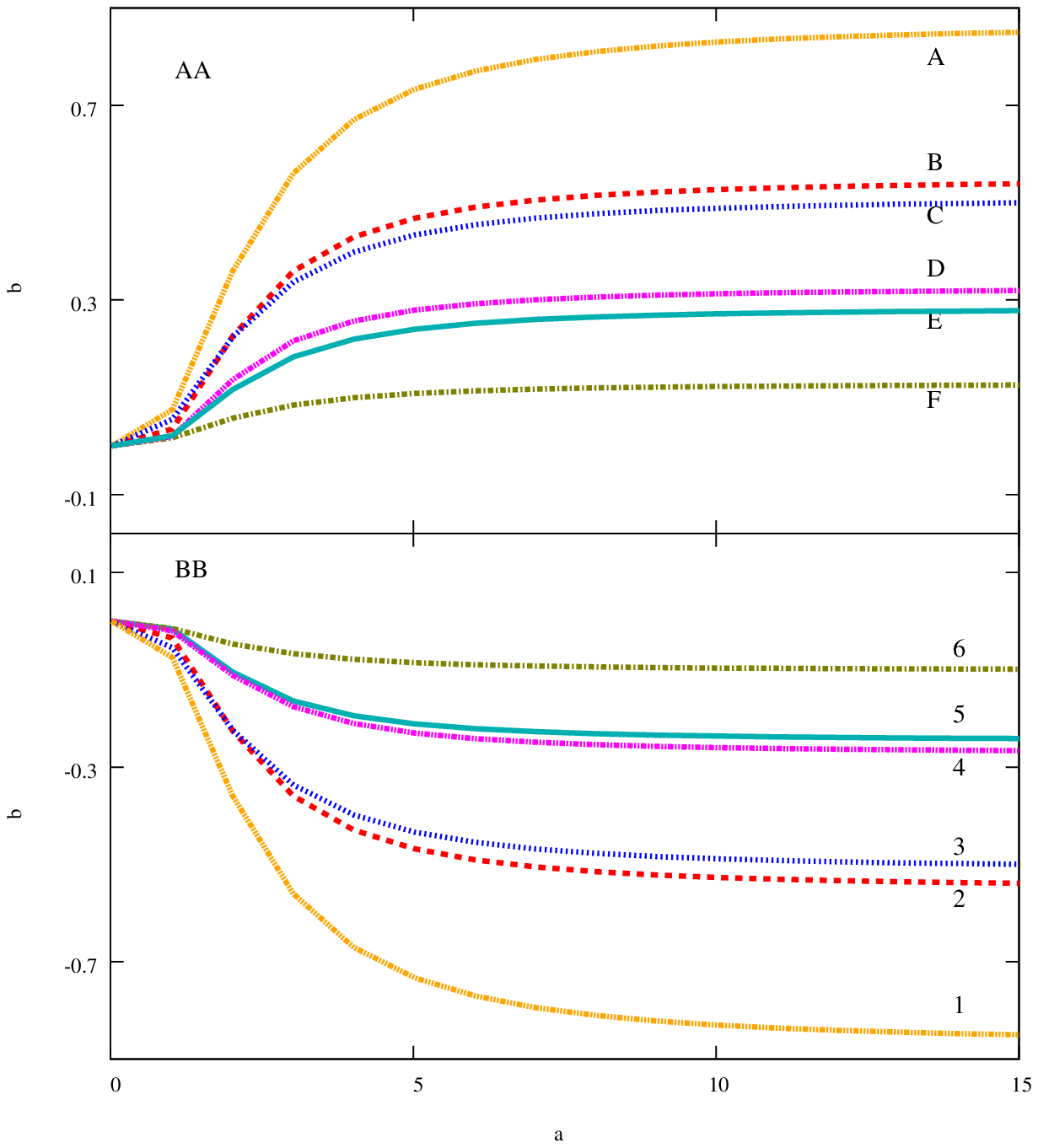}
  \caption{(I) Variation of $\kappa_{xy}(T)$ with $T$ for 
(a) $K_x=-0.8$, $K_y=-0.6$, $K^\prime_x=-1$, $K^\prime_y=0.4$ for ${\mathcal C}=(01\bar 10)$, 
(b) $K_x=-0.7$, $K_y=-0.2$, $K^\prime_x=-0.5$, $K^\prime_y=0.1$ for ${\mathcal C}=(100\bar 1)$, 
(c) $K_x=-0.8$, $K^\prime_x=-1$, $K^\prime_y=-0.4$ for ${\mathcal C}=(2\bar 33\bar 2)$,
(d) $K_x=-1$, $K_y=-0.1$, $K^\prime_x=-0.6$, $K^\prime_y=-0.3$ for ${\mathcal C}=(1\bar 22\bar 1)$, 
(e) $K_x=-0.7$, $K_y=-0.8$, $K^\prime_x=-0.1$, $K^\prime_y=-0.5$ for 
${\mathcal C}=(1\bar 11\bar 1)$, and
(f) $K_x=-0.7$, $K_y=-0.4$, $K^\prime_x=-0.8$, $K^\prime_y=-0.9$ for ${\mathcal C}=(2\bar 22\bar 2)$, with $K_z=-2$, $K^\prime_z=-1$, $D_{\rm m}=1$, $h=1$. (II) Variation of $\kappa_{xy}(T)$ for the conjugate phases. 
No value is assigned to those parameters when they are zero.}
   \label{Hallnew2}
  \end{figure}

However, in this study, a dozen of TMI phases is found 
in the presence of anisotropic Kitaev couplings on the NN and NNN bonds 
together with NNN DMI and external magnetic field. 
Band structures for six different TMI phases, are shown in Fig \ref{energy3dnew}.
Dispersion relations along the 
paths in BZ in addition to the DOS are 
shown in Fig \ref{energybznew}, for six different cases. 
 Dispersions of bulk-edge states in the one-dimensional BZ 
 have been shown in Fig \ref{edgenew}.  
A comprehensive topological phase diagram of the system 
for the anisotropic case is shown in Fig \ref{chern_no_new}. 
Variation of $\kappa_{xy}$ with respect to $K_y$ and $K'_y$ 
for different TMI phases is shown in Fig \ref{Hallnew1}.  
Those are plotted for $k_{B}T=20$. 
Similarly, variation of $\kappa_{xy}$ with respect to $T$ 
is shown in Fig \ref{Hallnew2}.    
All the figures are drawn 
for fixed values of the parameters,  
 $K_z=-2$, $K'_z=-1$, $D_{\rm m}=1$ and $h=1$. 

The remaining six TMI phases are defined by simultaneously reversing the sign of 
${\mathcal C}$ for all four bands, which are henceforth termed as 
the conjugate phases. Those conjugate phases are obtained by  
reversing the sign of either $K_y'$ alone or that of both 
$K_y$ and $K_y'$ simultaneously in the parameter space, 
but, without changing the signs and values of remaining other parameters. 
The specific values of $K_y$ and $K_y'$, in addition, will determine which criterion 
will be obeyed for the emergence of a particular conjugate phase. 
However, no figure corresponding to those conjugate phases is shown in this article. 

The band structure for TMI phase having ${\mathcal C}=(01\bar 10)$ 
is shown in Fig \ref{edgenew} (a), which is obtained for
 $K_x=-0.8$, $K_y=-0.6$, $K'_x=-1$ and $K'_y=0.4$. 
The conjugate TMI phase with ${\mathcal C}=(0\bar 110)$ 
appears if the value of $K'_y$ is changed to $-0.4$. 
System exhibits another TMI phase with ${\mathcal C}=(2\bar 33\bar 2)$ when NN Kitaev 
interactions along the $y$ bond is made zero, ($K_y=0$), 
keeping other parameters unchanged. 
The corresponding band structure is shown in Fig \ref{edgenew} (c). 
The conjugate TMI with ${\mathcal C}=(\bar 23\bar 32)$ is 
found just reversing the sign of $K'_y$.
TMI phase with ${\mathcal C}=(100\bar 1)$ emerges when $K_x=-0.7$, $K_y=-0.2$, 
$K'_x=-0.5$ and $K'_y=0.1$, which is shown in Fig \ref{edgenew} (b). 
But the conjugate TMI phase with ${\mathcal C}=(\bar 1001)$ 
appears in this case if both $K_y$ and $K'_y$ reverse their sign.  
The band structure of the system obtained for $K_x=-1$, $K_y=-0.1$, $K'_x=-0.6$ 
and $K'_y=-0.3$, is shown in Fig \ref{edgenew} (d). 
This corresponds to the topological phase having ${\mathcal C}=(1\bar 22\bar 1)$.
Conjugate of this TMI phase appears if $K'_y$ picks up the reverse sign.  
At $K_x=-0.7$, $K_y=-0.8$, $K'_x=-0.1$, TMI phase with 
${\mathcal C}=(1\bar 11\bar 1)$ and its conjugate ${\mathcal C}=(\bar 11\bar 11)$ 
appear for $K'_y=-0.5$, and $K'_y=0.5$, respectively. 
The former is shown in Fig \ref{edgenew} (e).
Finally, Fig \ref{edgenew} (f) corresponds to the band structure of 
another TMI phase with ${\mathcal C}=(2\bar 22\bar 2)$. 
The corresponding conjugate TMI phase, ${\mathcal C}=(\bar 22\bar 22)$, appears  
when signs of both $K_y$ and $K'_y$ are reversed.
The TMI phase, ${\mathcal C}=(1\bar 11\bar 1)$, is found to 
appear in both the cases of isotropic and anisotropic Kitaev couplings. However,
no conjugate phase is found in the isotropic case.
  
A comprehensive topological phase diagram of the system for the anisotropic case 
is shown in Fig \ref{chern_no_new}
for $K_x=-0.7$. Two diagrams (a) and (b) are drawn by varying $K_y$ and $K_y^\prime$, respectively, 
where the value of $K_x^\prime$ remains fixed at $-0.8$ and $-0.6$ in the respective cases. 
Trivial regions are indicated by $(0000)$, 
where all the ${\mathcal C}$s are zero.  
    
Variation of $\kappa_{xy} \hbar/k_{B}$ in the parameter space 
is shown in Fig \ref{Hallnew1} when 
$K_x=-0.7$, $K_z=-2$, $K'_z=-1$, $D_{\rm m}=1$, $h=1$, and $k_B T=20$. 
Here, different regions are identified with distinct colors,
those are used before in Fig \ref{chern_no_new}. 
Sudden jump in $\kappa_{xy}$ is noted where TPT takes place. 
Variation of $\kappa_{xy}$ with respect to $T$ 
for six different TMI phases along with their 
conjugate phases is shown in Fig \ref{Hallnew2}, with different colors. 
Signs of $\kappa_{xy}$ for a particular TMI phase 
and its conjugate are of opposite to each other. This corresponds to the 
fact that signs of the ${\mathcal C}$s of a definite phase 
are opposite to those of the corresponding conjugate phase. 
Thus, $\kappa_{xy}$ for all the twelve distinct topological phases have been 
shown in this figure.

\section{Discussion}
\label{Discussion}
Topological properties based on the bosonic magnon excitation of the FM  
 Kitaev-Heisenberg model on CaVO lattice in the presence of 
DMI have been investigated extensively in 
this study. The model comprises of Heisenberg and 
Kitaev terms both on NN and NNN bonds. Both isotropic as well as anisotropic 
couplings are considered. A sizable number of TMI phase 
appears upon variation of parameter values. Topological phases 
have been characterized in terms of Chern numbers which are evaluated 
numerically. DMI on a particular combination of NNN bonds 
is found crucial for the emergence of the TMIs. 
On the other hand, NN DMI has no role for the same. 

It has been indicated in the previous 
studies that FM Kitaev model with 
NN anisotropic couplings exhibits a unique topological phase based on the 
Majorana fermion representation both for the honeycomb and 
CaVO lattices in the presence of magnetic field \cite{Kitaev,Sun}. 
Interestingly, the situation is different in case of FM 
Kitaev-Heisenberg system, however, when solved in terms of 
bosonic magnon excitations. 
The present investigation reveals that 
multiple topological phases arise in Kitaev-Heisenberg model 
on CaVO lattice when both NN and NNN terms are there. 
No topological phase is there for NN term alone.
In addition, topological properties of the Kitaev-Heisenberg 
model on the honeycomb lattice are drastically different from 
those of CaVO lattice perhaps because of their 
different symmetries. 

In case of honeycomb lattice, nontriviality in the two-band 
Kitaev-Heisenberg system is induced by the presence of 
SAI term, $\Gamma\,(S^\alpha_i S^\beta_j\!
+\!S^\beta_i S^\alpha_j)$ \cite{Moumita}. Conjugate topological phases appear upon 
sign reversal of $\Gamma$, the strength of SAI. 
But no role of DMI is found there. 
On the other hand, no effect of SAI term is found 
on the topological properties of four-band 
Kitaev-Heisenberg system for CaVO lattice, 
where, NNN DMI term is found indispensable. 
Conjugate phases appear when the signs of Kitaev terms, 
$K_y$ and $K'_y$ are reversed depending on the situation. No conjugate phase appears in the fermionic
Kitaev models.

NNN Kitaev and Heisenberg terms could not lead to new topological 
phase for the honeycomb lattice \cite{Moumita}. But a number of 
new topological phases emerge as soon as the third neighbor 
Kitaev and Heisenberg terms are introduced. On the other hand, 
the system hosts multiple topological phases 
in the presence of NNN Kitaev and Heisenberg terms, in case of 
CaVO lattice.   
It is expected that numerous novel topological phases with 
higher values of Chern numbers will come up if third neighbor 
Kitaev and Heisenberg terms are taken into account. 
Therefore, the topological properties of FM Kitaev-Heisenberg models 
on both CaVO and honeycomb lattices are different 
when studied in terms of bosonic magnon excitations in comparison to the 
exactly solvable FM Kitaev models on the same lattices based on 
the Majorana representation. 
  \section{ACKNOWLEDGMENTS}
MD acknowledges the UGC fellowship, no. 524067 (2014), India.

\end{document}